\documentclass[12pt, a4paper]{article}
\usepackage{graphicx}
\usepackage{amssymb}
\usepackage{amsmath}
\usepackage{bm}
\usepackage{color}
\usepackage{theorem}
\usepackage{colortbl}
\usepackage{longtable}
\usepackage{subcaption}

\usepackage[sort&compress,numbers, merge]{natbib}

\definecolor{Orange}{cmyk}{0,0.61,0.87,0}
\definecolor{LightSalmon}{cmyk}{0,0.25,0.3,0}
\definecolor{JungleGreen}{cmyk}{0.99,0,0.52,0}
\definecolor{OliveGreen}{cmyk}{0.64,0,0.95,0.40}
\definecolor{Brown}{cmyk}{0,0.70,1,0.40}
\definecolor{RoyalBlue}{cmyk}{0.71,0.53,0,0.12}
\definecolor{Gray}{cmyk}{0,0,0,0.40}
\definecolor{LightPink}{cmyk}{0.0,0.25,0,0}
\definecolor{LLightPink}{cmyk}{0.0,0.10,0,0}
\definecolor{LightBlue}{cmyk}{0.25,0,0,0}
\definecolor{LightGray}{cmyk}{0,0,0,0.2}
\definecolor{LightGreen}{cmyk}{0.1,0,0.15,0}

\allowdisplaybreaks[1]

\usepackage{hyperref}

\begin{document}

\begin{titlepage}

\begin{flushright}
FTPI--MINN--15/47 \\
UMN--TH--3510/15\\
KIAS-P15063
\end{flushright}

\vskip 1.35cm
\begin{center}

%\textcolor{RoyalBlue}
{\LARGE
{\bf
The ATLAS Diboson Resonance in \\[7pt]
Non-Supersymmetric SO(10)
}
}

\vskip 1.2cm

Jason L. Evans$^{1,2}$,
Natsumi Nagata$^2$,
Keith A. Olive$^2$,
and
Jiaming Zheng$^2$

\vskip 0.5cm

{\it $^1${School of Physics, KIAS, Seoul 130-722, Korea}\\
 \vspace{0.25cm}
$^2${William I. Fine Theoretical Physics Institute, School of
 Physics and Astronomy, University of Minnesota, Minneapolis, Minnesota 55455,
 USA}
 }

\date{\today}

\vskip 1.5cm

\begin{abstract}

 SO(10) grand unification accommodates intermediate gauge symmetries
 with which gauge coupling unification can be realized without
 supersymmetry. In this paper, we discuss the possibility that a new
 massive gauge boson associated with an intermediate gauge symmetry
 explains the excess observed in the diboson resonance search recently
 reported by the ATLAS experiment. The model we find has two intermediate
 symmetries, $\text{SU}(4)_C \otimes \text{SU}(2)_L \otimes \text{SU}(2)_R$
 and $\text{SU}(3)_C \otimes \text{SU}(2)_L \otimes \text{SU}(2)_R
 \otimes \text{U}(1)_{B-L}$, where the latter gauge group is broken at
 the TeV scale. This model achieves gauge coupling unification
 with a unification scale sufficiently high to avoid proton decay. In addition, this model
 provides a good dark matter candidates, whose stability is guaranteed by a
 $\mathbb{Z}_2$ symmetry present after the spontaneous breaking of the
 intermediate gauge symmetries. We also discuss prospects for testing
 these models in the forthcoming LHC experiments and dark matter
 detection experiments.

\end{abstract}

\end{center}
\end{titlepage}

%%%%%%%%%%%%%%%%%%%%%%%%%%%%%%%%%%%%%%%%%%
\section{Introduction}
%%%%%%%%%%%%%%%%%%%%%%%%%%%%%%%%%%%%%%%%%%

Grand unification \cite{Georgi:1974sy} is considered to be one of the most
promising frameworks for physics beyond the Standard Model
(SM). SO(10) Grand Unified Theories (GUTs)
\cite{Georgi:1974my,so10-2,GN2} have many especially attractive features. First, the full
set of SM fermions of each generation together with a right-handed neutrino is
embedded into a  single ${\bf 16}$ chiral representation of SO(10). Second,
anomaly cancellation in the SM can be understood naturally since SO(10) is
free from anomalies. Third, gauge coupling unification can be realized without relying on supersymmetry, if there are intermediate-scale
gauge symmetries \cite{Rajpoot:1980xy}. Majorana mass terms of
right-handed neutrinos can be generated at an intermediate scale,
which may explain the smallness of neutrino masses. These interesting aspects
have stimulated many extensive studies of SO(10) GUTs
\cite{Fukugita:1993fr, DiLuzio:2011my, Fukuyama:2012rw, so10-3}.

In addition, SO(10) GUTs with intermediate gauge symmetries predict new gauge bosons
whose masses are of the order of the breaking scales of the corresponding
gauge symmetries. The breaking scale of these intermediate gauge symmetries affects the running of the gauge couplings. These scales can be determined by requiring gauge
coupling unification, which is highly dependent on the symmetry-breaking pattern
and matter content of the model. However, we will show that it is possible to
obtain a model containing TeV-scale gauge bosons accessible at the LHC,  for a particular choice
of the intermediate gauge symmetries and particle content.

In this paper, we consider non-supersymmetric SO(10) GUT models with an
intermediate gauge symmetry showing up around the TeV scale, and discuss
the possibility that a new gauge boson associated with the intermediate gauge symmetry can
explain the anomalous diboson events recently reported by the ATLAS
collaboration \cite{Aad:2015owa}. We will discuss a model with two
intermediate symmetries, $\text{SU}(4)_C \otimes \text{SU}(2)_L \otimes
\text{SU}(2)_R$ which is then broken to $\text{SU}(3)_C \otimes \text{SU}(2)_L \otimes
\text{SU}(2)_R \otimes \text{U}(1)_{B-L}$.
The unified and intermediate scales are determined by the renormalization group (RG)
running of the gauge couplings and depend on the matter content of the theory.
In the model discussed,  the GUT scale is of order $10^{17}$ GeV and is
high enough so that the proton decay bounds can be evaded.
The first intermediate scale is of order $10^{9}$ GeV and is high enough that
constraints from leptoquarks are satisfied.
The second intermediate
scale lies around a few TeV and
the charged gauge bosons of $\text{SU}(2)_R$, $W_R^\pm$, can have a mass $\sim 2$~TeV and may
explain the excess observed in the ATLAS diboson resonance search
\cite{Aad:2015owa}. The possibility of this new charged gauge boson
explaining the diboson anomaly is discussed in
Refs.~\cite{Hisano:2015gna, Cheung:2015nha, Gao:2015irw,
Brehmer:2015cia, Dev:2015pga, Deppisch:2015cua}. We will see that $W_R^\pm$ in this model
can also reproduce the diboson excess. Furthermore, this $W_R^\pm$ may explain another anomaly observed by the CMS collaboration in the
right-handed neutrino searches with a dijet and a dilepton final state
\cite{Khachatryan:2014dka}, as we will discuss below. Our model also predicts an additional massive boson: a neutral gauge
boson $Z_R$. Detection of this $Z_R$ would be a smoking-gun signature of our
model. The masses of the $Z_R$ can be heavier than the $W_R^\pm$
mass, and thus evades the current LHC bound.
However, it may be in reach of Run-II of the LHC, as we will discuss.

Our model also contains a promising dark matter
candidate. The dark matter candidate is stabilized by a residual $\mathbb{Z}_2$
symmetry arising after the SO(10) symmetry has been completely broken to
the SM gauge symmetries \cite{Kibble:1982ae, Krauss:1988zc, Ibanez:1991hv,
Martin:1992mq,Kadastik:2009dj,Frigerio:2009wf, Mambrini:2015vna,
Nagata:2015dma, Arbelaez:2015ila, Boucenna:2015sdg}.
This residual $\mathbb{Z}_2$, as it turns out, is equivalent to matter parity
\cite{Farrar:1978xj}. It is tantalizing that the presence of this dark
matter candidate is essential for the model to achieve gauge coupling
unification. An integral part of obtaining gauge coupling unification in
this set up is an intermediate scale of order a TeV, a second intermediate scale
large enough to suppress the effects of leptoquarks, and a GUT scale
which is large enough to avoid proton decay.\footnote{Realization of
a few TeV intermediate scale within the framework of grand unification
is also discussed in Refs.~\cite{Aydemir:2015nfa, Bandyopadhyay:2015fka,
Aydemir:2015oob}.} Given these conditions, the
required matter content at low energies is severely restricted, and thus
it is highly non-trivial that our model contains a good
dark matter candidate. The presence of dark matter gives us another
opportunity to test this model; we can probe the dark matter candidate
predicted in this model in dark matter detection experiments.
We also discuss this prospect in what follows. Combining dark matter
search results with those obtained at the LHC experiments, we may be
able to test our model in near future.

This paper is organized as follows. We first describe our model in the
next section. Then, in Sec.~\ref{sec:dibosonsignal}, we briefly review the
ATLAS diboson anomaly, and show that the $W^\pm_R$ in our model can
explain the diboson anomaly. We also
discuss the present constraints on our models from various experiments
and consider the testability of this model in future LHC experiments
and dark matter searches. Finally, we conclude in Sec.~\ref{sec:conclusion}.

%%%%%%%%%%%%%%%%%%%%%%%%%%%%%%%%%%%%%%%%%%%
\section{The Model}
\label{sec:models}
%%%%%%%%%%%%%%%%%%%%%%%%%%%%%%%%%%%%%%%%

In this section, we describe the specific SO(10) model used in this analysis. This
model is obtained in a similar manner to that discussed in
Ref.~\cite{Nagata:2015dma}, except now we allow for two intermediate scales.
To obtain these intermediate
scales, we need to arrange some Higgs fields to acquire vacuum expectation
values (VEVs) in specific directions at particular scales. Throughout
this work, we make an assumption on the Higgs field content. We assume that at each
intermediate scale the Higgs fields present are components of the SO(10)
multiplets that have VEVs equal to or less then the corresponding intermediate scale. The
other components have masses of order of the
previous symmetry-breaking scale. We also allow for a light SO(10) component which
could be dark matter.

Before we get into the details of our model, we first discuss the
necessity for two intermediate scales. If there were only one
intermediate scale and we wished to explain the ATLAS diboson excess,
we would have to choose the TeV scale. For intermediate gauge symmetries which include the SU(4)$_C$ subgroup,
this choice turns out to be problematic since the breaking of the
SU(4)$_C$ subgroup generates a vector leptoquark at that intermediate scale.
Since this leptoquark
interacts with both left- and right-handed fermions, it will generate
excessively large contributions to the meson decays (see, \textit{e.g.},
Ref.~\cite{Davidson:1993qk}).  In fact, one can show that no tuning can simultaneously
tune away the contribution of a vector leptoquark to the Kaon decays
$K^+\to \pi^+ \bar e \mu$ and $K_L\to \bar e \mu$ by exploiting the
contributions of the scalar leptoquarks
contained inside the Higgs fields.\footnote{This is
because both of these decays arise from the same effective
operator. These two decays depend on the vector current and axial
current respectively.  Because the effective operators generated from
scalar leptoquarks will have different chirality for the quarks than the
vector leptoquark, the contribution from the scalar leptoquark will add
to one decay and subtract from the others.  Therefore, no tuning can
remove this contribution.}
If the intermediate gauge symmetries do not contain the SU(4)$_C$
subgroup, on the other hand, we find that we cannot obtain gauge
coupling unification with a sufficiently high GUT scale since
there are no new contributions to the running of the SU(3)$_C$ gauge
coupling constant. This makes it impossible to get the intermediate
scale to be around the TeV scale while maintaining gauge coupling
unification with a sufficiently high GUT scale. These problems, however, can be
evaded if a two step breaking pattern is considered. The vector
leptoquark  then has a mass of order of the larger intermediate scale. This larger intermediate scale generates a mass for the vector leptoquarks which is high enough to avoid constraints coming from meson decays but low enough that the vector leptoquarks still contribute significantly to the running of the SU(3)$_C$ gauge coupling.

The breaking pattern of our model, with two intermediate gauge symmetries, is from SO(10) to
$\text{SU}(4)_C \otimes \text{SU}(2)_L \otimes \text{SU}(2)_R$ which is then broken to
$\text{SU}(3)_C \otimes \text{SU}(2)_L \otimes \text{SU}(2)_R \otimes
\text{U}(1)_{B-L}$. For the breaking pattern we consider, SO(10) is broken by the VEV
of the $({\bf 1}, {\bf 1}, {\bf 1})$ component of a {\bf 210} into
$\text{SU}(4)_C \otimes \text{SU}(2)_L \otimes \text{SU}(2)_R$. This symmetry is then
subsequently broken by a VEV of the $({\bf 15}, {\bf 1}, {\bf 1})$
component of the same {\bf 210} to $\text{SU}(3)_C \otimes \text{SU}(2)_L
\otimes \text{SU}(2)_R \otimes \text{U}(1)_{B-L}$. Importantly, the $({\bf 1}, {\bf
1}, {\bf 1})$ component of the {\bf 210} breaks $D$-parity
\cite{Kuzmin:1980yp} at the GUT scale, and thus the low-energy effective
theory does not exhibit a left-right symmetry
\cite{Pati:1974yy}. Finally, this gauge group is broken to the SM gauge
symmetries at the TeV scale by the $({\bf 1}, {\bf 1}, {\bf 3}, +2)$ component
of a ${\bf 126}$. However, there remains a $\mathbb{Z}_2$ symmetry coming from
U(1)$_{B-L}$ since the $B-L$ charge of
this component of the ${\bf 126}$ is $2$ \cite{Kibble:1982ae, Krauss:1988zc, Ibanez:1991hv, Martin:1992mq}.  The $({\bf 1}, {\bf 1},
{\bf 3}, 0)$ component of the ${\bf 210}$ also breaks $\text{SU}(3)_C
\otimes \text{SU}(2)_L \otimes \text{SU}(2)_R \otimes
\text{U}(1)_{B-L}$, but does not break the $\mathbb{Z}_2$ symmetry since
its $B-L$ charge is zero.  In summary, we consider the following symmetry-breaking chain:
\begin{align}
 \text{SO}(10)& \stackrel{\langle {\bf 210} \rangle}{\longrightarrow}
\text{SU}(4)_C \otimes \text{SU}(2)_L \otimes
\text{SU}(2)_R
\nonumber \\
& \stackrel{\langle {\bf 210} \rangle}{\longrightarrow}
\text{SU}(3)_C \otimes \text{SU}(2)_L \otimes
\text{SU}(2)_R \otimes \text{U}(1)_{B-L}
\nonumber \\
&\xrightarrow{\langle {\bf 126} \rangle~ \langle {\bf 210} \rangle} ~
\text{SU}(3)_C \otimes \text{SU}(2)_L \otimes \text{U}(1)_Y
\otimes \mathbb{Z}_2 ~.
\label{eq:symmetrychain2}
\end{align}
Under the remnant $\mathbb{Z}_2$ symmetry, which is found to be
equivalent to matter parity $P_M =(-1)^{3(B-L)}$
\cite{Farrar:1978xj}, the SM fermions are odd while the SM Higgs is
even. Therefore, a scalar boson (fermion) can be stable if it is odd (even)
under this $\mathbb{Z}_2$ symmetry. If such a particle is electrically
and color neutral, it can be a good dark matter candidate
\cite{Kadastik:2009dj,Frigerio:2009wf}. For a detailed
discussions on this class of dark matter candidates in SO(10) GUTs, see
Refs.~\cite{Mambrini:2015vna, Nagata:2015dma}.

The low-energy matter content of this model, beyond the
SM fermions and right-handed neutrinos (which are embedded into three
{\bf 16} representations), is given in Table~\ref{tab:model2}. Here, the first, second, and third
columns show the SO(10), $\text{SU}(4)_C \otimes \text{SU}(2)_L \otimes
\text{SU}(2)_R$, and $\text{SU}(3)_C \otimes \text{SU}(2)_L \otimes
\text{SU}(2)_R \otimes \text{U}(1)_{B-L}$ quantum numbers of the
particles, respectively, while the fourth column represents their
$\mathbb{Z}_2$ charges. The subscripts $C$ and $D$ refer to complex and Dirac respectively.
Most of the components of the {\bf 210} have GUT scale masses and do not affect our
analysis.
In this model, the ({\bf 15}, {\bf 1}, {\bf
1})$_C$ component of {\bf 210}$_C$ has a mass similar to the $\text{SU}(4)_C
\otimes \text{SU}(2)_L \otimes \text{SU}(2)_R$ breaking scale, as
its VEV breaks the symmetry.
Similarly, the components of({\bf 15}, {\bf 1}, {\bf 3}) $\in$  {\bf 210} and
($\overline{\bf 10}$, {\bf 1}, {\bf 3}) $\in$ {\bf 126} charged under color have masses of order the higher intermediate scale breaking. The components in the
third column of the table are those components which are fine-tuned to have masses lighter than
this scale. Among the light fields are the ({\bf 1}, {\bf 1},
{\bf 3}, $+2$)$_C$ and the ({\bf 1}, {\bf 1}, {\bf 3}, 0)$_C$ which obtain VEVs and break the $\text{SU}(3)_C \otimes \text{SU}(2)_L \otimes
\text{SU}(2)_R \otimes \text{U}(1)_{B-L}$ symmetry to the SM symmetries. The components of these fields which are not eaten by the gauge bosons receive masses of order of the VEV breaking this symmetry.  The rest of
the Higgs fields listed in the table are responsible for breaking the electroweak
symmetry and have masses of order the electroweak scale. Also listed in the table is an $\text{SU}(2)_L$
triplet Dirac fermion which can be a viable dark matter candidate and could originate from a {\bf 45} of SO(10).
The remaining components of the {\bf 45} (beyond the triplet) should all have GUT scale masses.
In summary, the states in column 2 affect the running of
the gauge couplings between the high intermediate scale
and the GUT scale, and those in column 3 affect the running between the TeV scale and the
high intermediate scale. The particles in the third column and green shaded rows contribute to the running of the gauge couplings below the TeV scale. This leads to a theory below the TeV scale with the SM fermions and four doublet Higgs bosons\footnote{As we will see below, the dark matter candidate will need to be of order the TeV scale.}. 
A similar decomposition can be made for the the
gauge bosons; the $({\bf 6}, {\bf 2}, {\bf 2})$ component of the SO(10)
gauge boson, ${\bf 45}$, is around the GUT scale, while the other
components, the SU(4)$_C$, SU(2)$_L$, and SU(2)$_R$ gauge bosons, are
massless at this stage of the GUT symmetry breaking. A part of the
SU(4)$_C$ gauge bosons, the vector leptoquark, obtains a mass of the order
of the
higher intermediate scale, as we will see below.

%%%%%%%%%%%%%%%%%%%%% TABLE %%%%%%%%%%%%%%%%%%%%%%%%%%%%%%%%%%%%%%%%%%
\begin{table}[h]
 \begin{center}
\caption{\it Particle content of our model. The first, second, and third
  columns show the SO(10), $\text{SU}(4)_C \otimes \text{SU}(2)_L \otimes
  \text{SU}(2)_R$, and $\text{SU}(3)_C \otimes \text{SU}(2)_L \otimes
  \text{SU}(2)_R \otimes \text{U}(1)_{B-L}$ quantum numbers of the
  particles, respectively, while the forth column represents their
  $\mathbb{Z}_2$ charges. The orange (green) shaded fields acquire VEVs
  of about a few TeV (the electroweak scale), and the blue shaded
  particle contains a dark matter candidate. The $({\bf 15}, {\bf 1}, {\bf
  1})_C$ of the {\bf 210}$_C$ lies around the $\text{SU}(4)_C
  \otimes \text{SU}(2)_L \otimes \text{SU}(2)_R$ breaking scale, and none of
  its components appears below this scale. }
\label{tab:model2}
\vspace{5pt}
\begin{tabular}{cccc}
\hline
\hline
SO(10) &
$\text{SU}(4)_C \otimes \text{SU}(2)_L \otimes \text{SU}(2)_R$
& $\text{SU}(3)_C \otimes \text{SU}(2)_L \otimes \text{SU}(2)_R \otimes
\text{U}(1)_{B-L}$
&  $\mathbb{Z}_2$\\
\hline
{\bf 210}$_C$ & ({\bf 15}, {\bf 1}, {\bf 1})$_C$ & -- &$+$\\
\rowcolor{LightSalmon}
{\bf 126}$_C$ & ($\overline{\bf 10}$, {\bf 1}, {\bf 3})$_C$
& ({\bf 1}, {\bf 1}, {\bf 3}, $+2$)$_C$& $+$\\
\rowcolor{LightSalmon}
{\bf 210}$_C$ &({\bf 15}, {\bf 1}, {\bf 3})$_C$ &
({\bf 1}, {\bf 1}, {\bf 3}, 0)$_C$ &$+$\\
\rowcolor{LightGreen}
{\bf 10}$_C$ &({\bf 1}, {\bf 2}, {\bf 2})$_C$
& ({\bf 1}, {\bf 2}, {\bf 2}, 0)$_C$&  $+$\\
\rowcolor{LightGreen}
{\bf 126}$_C$ &({\bf 15}, {\bf 2}, {\bf 2})$_C$
& ({\bf 1}, {\bf 2}, {\bf 2}, 0)$_C$&  $+$\\
\rowcolor{LightBlue}
{\bf 45}$_D$ &({\bf 1}, {\bf 3}, {\bf 1})$_D$ &({\bf 1}, {\bf 3}, {\bf
	 1}, 0)$_D$&  $+$\\
\hline
\hline
\end{tabular}
 \end{center}
\end{table}
%%%%%%%%%%%%%%%%%%%%%%%%%%%%%%%%%%%%%%%%%%%%%%%%%%%%%%%%%%%%%%%%%%%%%%%

The mass of the dark matter candidate is fixed by requiring that its thermal relic
abundance agree with the observed dark matter density
$\Omega_{\text{DM}} h^2 \simeq 0.12$ \cite{Ade:2015xua}. The thermal
relic abundance of any DM candidate is determined by its thermally averaged annihilation
cross section $\sigma_{\text{ann}}$ times the relative velocity between
the initial particles, $v_{\text{rel}}$, via the following relation:
\begin{equation}
\Omega_{\text{DM}} h^2 \simeq
\frac{3 \times 10^{-27}~\text{cm}^3 ~\text{s}^{-1}}{\langle
 \sigma_{\text{ann}} v_{\text{rel}} \rangle} ~.
\label{eq:DMab}
\end{equation}
Here, the brackets signify a thermal average. In the case of an SU(2)$_L$
triplet Dirac fermion dark matter candidate with zero hypercharge, the $s$-wave
annihilation cross section is given by \cite{Cirelli:2005uq}
\begin{equation}
 \langle \sigma_{\text{ann}} v_{\text{rel}}\rangle
\simeq
\frac{37 g_{2}^4}{192 \pi m^2_{\text{DM}}} ~,
\label{eq:sigv}
\end{equation}
where $g_{2}$ denotes the SU(2)$_L$ gauge coupling constant and
$m_{\text{DM}}$ is the dark matter mass. From Eqs.~\eqref{eq:DMab} and
\eqref{eq:sigv}, we obtain $m_{\text{DM}} \simeq 2$~TeV. However, it
is known that the non-perturbative Sommerfeld effect significantly
enhances the annihilation cross section \cite{Hisano:2003ec}, which
results in a smaller cross section, and thus larger dark matter mass, needed to reproduce the correct dark
matter density. As far as we know, the thermal relic abundance of a
triplet Dirac fermion has not been computed yet with the
Sommerfeld enhancement effect included. However, we expect that the
favored mass for a Dirac triplet is smaller than that for a Majorana
triplet (which was found to be $\sim 2.7$~TeV
\cite{Hisano:2006nn}), since the thermally-averaged annihilation cross
section of Dirac dark matter is a factor of two smaller than that of
Majorana dark matter.

\begin{figure}[h]
\begin{center}
\includegraphics[height=80mm]{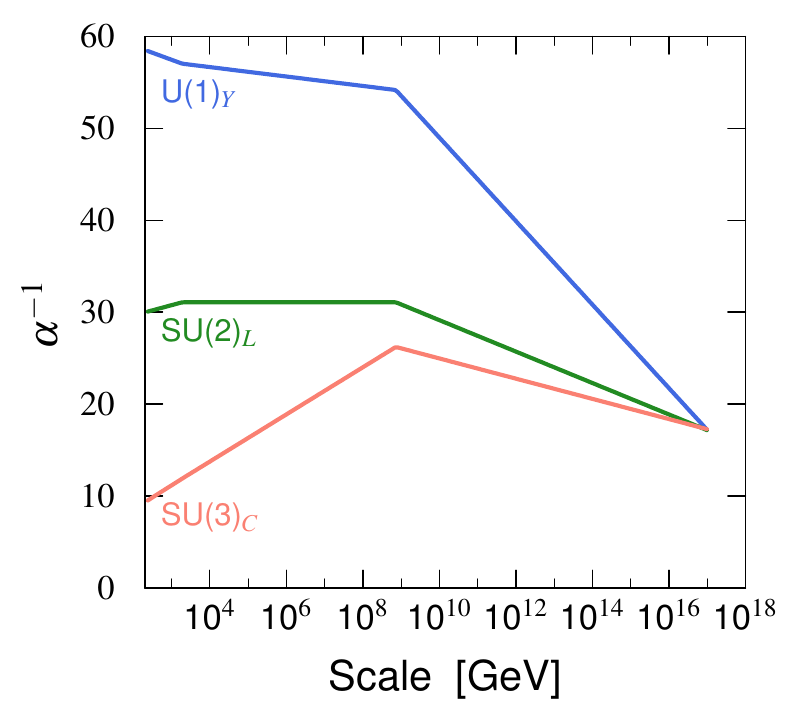}
\end{center}
%\vspace{0.3cm}
\caption{{\it Running of gauge couplings in our model. Here, we set
 the masses of the fields in the third column in Table~\ref{tab:model2}
 to be 1.9~TeV.}}
\label{fig:rge2}
\end{figure}

In Fig.~\ref{fig:rge2}, we show a plot of the running of the gauge coupling
constants in this model. Here, we set the masses of the
fields in the third column in Table~\ref{tab:model2} to be 1.9~TeV.
In this figure, we show the running of the inverse of $\alpha_a \equiv
g_a^2/ (4\pi)$ $(a = 1,2,3)$ with $g_1$, $g_2$, and $g_3$ corresponding
to the U(1)$_Y$, SU(2)$_L$, and SU(3)$_C$ gauge coupling constants,
respectively.
The definitions of these gauge couplings in the $\text{SU}(4)_C \otimes
\text{SU}(2)_L \otimes \text{SU}(2)_R$ theory are given by
\begin{equation}
 \frac{1}{\alpha_1} \equiv \frac{3}{5}\frac{1}{\alpha_{2R}} +
\frac{2}{5}\frac{1}{\alpha_4} ~, ~~~~
\alpha_2 \equiv \alpha_{2L}, ~~~~\alpha_3 \equiv \alpha_4 ~,
\label{eq:alp422}
\end{equation}
where $\alpha_4$, $\alpha_{2L}$, and $\alpha_{2R}$ represent the
SU(4)$_C$, SU(2)$_L$, and SU(2)$_R$ gauge couplings, respectively,
while those in the $\text{SU}(3)_C \otimes
\text{SU}(2)_L \otimes \text{SU}(2)_R \otimes \text{U}(1)_{B-L}$ theory are
\begin{equation}
 \frac{1}{\alpha_1} \equiv \frac{3}{5}\frac{1}{\alpha_{2R}} +
\frac{2}{5}\frac{1}{\alpha_{B-L}} ~, ~~~~
\alpha_2 \equiv \alpha_{2L},
\label{eq:alp3221}
\end{equation}
where $\alpha_{B-L}$ is the U(1)$_{B-L}$ gauge coupling. Using the weak scale values of the gauge couplings as inputs and insisting on
unification of the coupling constants at some high energy scale, allows us to use the three RGEs
to solve for the higher intermediate scale, the GUT scale, and the value of the
the unified gauge coupling at the GUT scale (recall that we have fixed the value of the lower
intermediate scale).
From this
figure, we find that the GUT scale in this model is $9.8 \times 10^{16}$~GeV, which is high enough to avoid
proton decay constraints. The $\text{SU}(4)_C \otimes \text{SU}(2)_L
\otimes \text{SU}(2)_R$ breaking scale is determined to be $\simeq
7.2\times 10^8$~GeV. The $\text{SU}(3)_C \otimes \text{SU}(2)_L \otimes
\text{SU}(2)_R \otimes \text{U}(1)_{B-L}$ gauge coupling constants at
1.9~TeV are found to be
\begin{equation}
 g_{B-L} \simeq 0.58~, ~~~~
 g_{2L} \simeq 0.64~, ~~~~
 g_{2R} \simeq 0.42~, ~~~~
 g_{3} \simeq 1.03~.
\label{eq:gaugcoup3221}
\end{equation}
The value of $g_{2R}$ above will be important when we discuss the diboson
anomaly.

%%%%%%%%%%%%%%%%%%% FIGURE %%%%%%%%%%%%%%%%%%%%%%%%%%%%%%%%%%%%%%%%%%%%%
%%%%%%%%%%%%%%%%%%%%%%%%%%%%%%%%%%%%%%%%%%%%%%%%%%%%%%%%%%%%%%%%%%%%%%%

In this model, only $W_R^\pm$ and $Z_R$ lie around the TeV scale, while the
vector leptoquarks have masses of ${\cal O}(10^{\text{8--9}})$~GeV. The scalar leptoquarks arising from the $({\bf 15}, {\bf 1}, {\bf 1})_C$
and $({\bf 15}, {\bf 1}, {\bf 3})_C$ of the ${\bf 210}_C$ and
$(\overline{\bf 10}, {\bf 1}, {\bf 3})_C$ and $({\bf 15}, {\bf 2}, {\bf
2})_C$ of the ${\bf 126}_C$ also have a mass of order this scale. The
masses of $W_R^\pm$ and $Z_R$ are computed to be
\begin{align}
 m_{W_R}^2 &\simeq g_{2R}^2 \left[
(v_R^{\bf 10})^2 + (v_R^{\bf 15})^2
\right] ~,
\nonumber \\
 m_{Z_R}^2 &\simeq
2\left(g_{2R}^2 +\frac{3}{2}g_{B-L}^2\right)(v_R^{\bf 10})^2 ~,\label{eqn:MZR}
\end{align}
where
\begin{equation}
 \langle ({\bf 1}, {\bf 1}, {\bf 3}, +2)_C \rangle = T_- v_R^{\bf 10} ~,
~~~~~~
 \langle ({\bf 1}, {\bf 1}, {\bf 3}, 0)_C \rangle = T_3 v_R^{\bf 15} ~.
\end{equation}
Here, $T_a = \tau_a/2$ ($a =1,2,3$) with $\tau_a$ the Pauli matrices and
$T_- \equiv T_1-iT_2$, and we have neglected the contribution of the
electroweak-scale VEVs for simplicity.
On the other hand, the vector leptoquarks, whose quantum numbers under
the $\text{SU}(3)_C \otimes \text{SU}(2)_L \otimes \text{U}(1)_Y$
symmetry are $({\bf 3}, {\bf 1}, +2/3) \oplus (\overline{\bf 3}, {\bf
1}, -2/3)$, acquire a mass of
\begin{equation}
 m_{V_{\text{LQ}}}^2 \simeq \frac{2}{3} g_4^2 \left(v_{422}^{\bf
15}\right)^2 ~,
\label{eq:mlq}
\end{equation}
with
\begin{equation}
 \langle ({\bf 15}, {\bf 1}, {\bf 1})_C \rangle = X_{B-L} v_{422}^{\bf
  15} ~,
\end{equation}
where $X_{B-L}$ is the SU(4)$_C$ generator corresponding to the $B-L$ charge, i.e. $\sqrt{\frac{3}{8}}(B-L)$.\footnote{As a simple cross check,
we here note that for a generic
gauge theory which is spontaneously broken by a VEV of a Higgs field
$\langle \varphi \rangle$, the trace of the mass matrix $M^2_{AB}$ of the gauge
bosons is given by
\begin{equation}
 \sum_{A}M^2_{AA} = 2 g_G^2 \langle \varphi \rangle^* T_AT_A \langle
  \varphi \rangle = 2g_G^2 C_G (R)|\langle \varphi \rangle|^2 ~,
\label{eq:genmaa}
\end{equation}
where $C_G(R)$ denotes the quadratic Casimir invariant for a
representation $R$ of a gauge group $G$. For example, the vector
leptoquark mass $m_{V_{\text{LQ}}}^2$ is derived from the relation as
\begin{equation}
 6m_{V_{\text{LQ}}}^2 = 2 g_4^2 C_{\text{SU}(4)} ({\bf 15}) \cdot
| \langle ({\bf 15}, {\bf 1}, {\bf 1})_C \rangle|^2 ~.
\end{equation}
For the TeV-scale gauge bosons, on the other hand, we have
\begin{align}
 2m_{W_R}^2 + m_{Z_R}^2 &= 2 g_{2R}^2 C_{\text{SU}(2)}({\bf 3})\cdot \left[
| \langle ({\bf 1}, {\bf 1}, {\bf 3}, +2)_C \rangle|^2 +
|\langle ({\bf 1}, {\bf 1}, {\bf 3}, 0)_C \rangle |^2
\right] \nonumber \\
&+ 2g_{B-L}^2 \left(2\sqrt{\frac{3}{8}}\right)^2 \cdot | \langle ({\bf 1}, {\bf 1}, {\bf 3}, +2)_C \rangle|^2 ~.
\end{align}
These two relations are consistent with Eqs.~\eqref{eqn:MZR} and
\eqref{eq:mlq}.
} As the $ ({\bf 15}, {\bf 1}, {\bf
1})_C$ VEV $v_{422}^{\bf 15}$ sets the $\text{SU}(4)_C \otimes
\text{SU}(2)_L \otimes \text{SU}(2)_L$ breaking scale,
$m_{V_{\text{LQ}}}$ is as high as $\sim 10^{9}$~GeV. The masses of the
scalar leptoquarks are dependent on the couplings in the scalar
potential; generically, they are also ${\cal O}(v^{\bf 15}_{422})$.

\begin{figure}[h]
\begin{center}
\includegraphics[height=80mm]{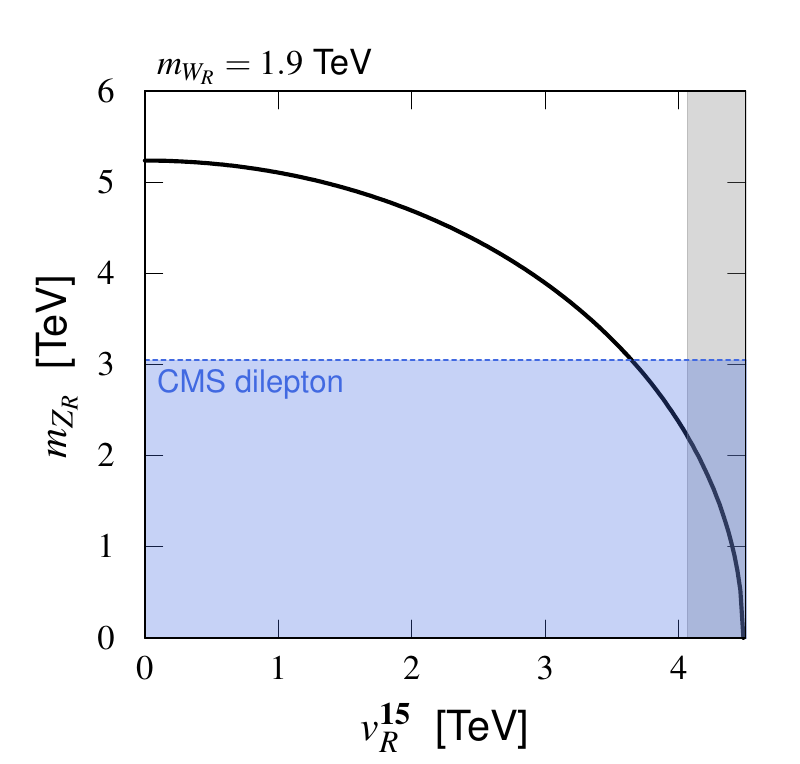}
\end{center}
%\vspace{0.3cm}
\caption{{\it $m_{Z_R}$ as a function of $v_R^{\bf 15}$. Here, we set
 $m_{W_R} = 1.9$~TeV. The gray shaded region predicts right-handed
 neutrino masses to be lighter than the $W_R^\pm$ mass when $y_\Delta
 <1$, while the blue shaded area is excluded by the CMS dilepton
 searches \cite{Khachatryan:2014fba}. }}
\label{fig:zmass2}
\end{figure}

Since the gauge couplings $g_{B-L}$ and $g_{2R}$ are determined, 
and given in Eq.~\eqref{eq:gaugcoup3221}, if we fix the $W_R^\pm$ mass, then we
obtain the $Z_R$ mass as a function of $v_R^{\bf 15}$. We show this in
Fig.~\ref{fig:zmass2} setting $m_{W_R} = 1.9$~TeV. It is found that
the $Z_R$ mass is relatively small; in particular, $m_{Z_R}$ can be much
smaller than that predicted in the minimal left-right symmetric
model with a single SU(2)$_R$ triplet Higgs field \cite{Pati:1974yy}, which is
reproduced if one takes $v_R^{\bf 15} = 0$. This distinguishing feature
can be tested at the LHC. Currently, the most stringent bound on
$m_{Z_R}$ is given by
the resonance searches in Drell-Yan processes \cite{Aad:2014cka,
Khachatryan:2014fba}. The CMS collaboration gives the most
severe limits on this process \cite{Khachatryan:2014fba}. Using this analysis\footnote{The CMS limits are given in terms of the two parameters $c_u$ and $c_d$, which are defined in \cite{Accomando:2010fz}, as
$c_{u,d} = (g'^2/2)({g_V^{u,d}}^2 +{g_A^{u,d}}^2)Br(\ell \ell)$, where $Br(\ell \ell)$ is the branching fraction into lepton pairs, and $g^\prime
g_V^{u (d)}$ and $g^\prime g_A^{u (d)}$ are the vector and axial-vector
couplings of $Z_R$ with up-type (down-type) quarks, respectively. Using these expressions we find that our model gives $c_u
\simeq 2\times 10^{-3}$ and $c_d \simeq 1\times 10^{-2}$, which leads to our stated mass bound on $Z_R$. }, it is found that
the present lower bound on the $Z_R$ mass in our model is $m_{Z_R}\gtrsim 3.05$~TeV\footnote{
This limit is relaxed slightly if one allows $Z_R$ to decay into right-handed neutrinos.}.
We show this limit by the blue shaded area in Fig.~\ref{fig:zmass2}.
We find that $m_{Z_R}$ in our model can satisfy this constraint over a
wide range of $v_R^{\bf 15}$. The rest of allowed
mass range, $3.05$~TeV $\lesssim m_{Z_R}\lesssim 5.2$~TeV, can be reached at Run-II of the LHC. For example, the 14~TeV LHC run with an integrated luminosity of
300~fb$^{-1}$ can probe the entire parameter space in
Fig.~\ref{fig:zmass2} \cite{Godfrey:2013eta}.
Finally, we note in passing that the $Z_R$ in our model may explain the
dielectron event at 2.9~TeV recently announced by the CMS collaboration
\cite{CMS-DP-2015-039}, although there are inconsistencies with the
CMS dilepton bound. These $Z_R$ signatures can possibly be tested in the near future.

%%%%%%%%%%%%%%%%%%% FIGURE %%%%%%%%%%%%%%%%%%%%%%%%%%%%%%%%%%%%%%%%%%%%%
%%%%%%%%%%%%%%%%%%%%%%%%%%%%%%%%%%%%%%%%%%%%%%%%%%%%%%%%%%%%%%%%%%%%%%%

The VEVs of the bi-doublet Higgs fields $\Phi_{\bf 1}$ and $\Phi_{\bf
15}$ originally coming from $({\bf 1}, {\bf 2}, {\bf
2})_C$ and $({\bf 15}, {\bf 2}, {\bf 2})_C$ of the ${\bf 10}_C$ and ${\bf
126}_C$, respectively, induce mixing between the
SU(2)$_L$ and SU(2)$_R$ gauge bosons. Through this mixing, $W_R^\pm$ can
decay into a $W$ and $Z$. Therefore, by appropriately choosing
the mixing angle, we may explain the diboson excess observed in the ATLAS
experiment \cite{Aad:2015owa}, as we discuss in the next section.

In addition, the VEVs of $\Phi_{\bf 1}$, $\Phi_{\bf 15}$, and $\Delta =
({\bf 1}, {\bf 1}, {\bf 3}, +2)_C$ from the ${\bf 126}_C$ generate the
fermion mass terms via Yukawa interactions. At the TeV scale, the
relevant Yukawa terms are given by
\begin{align}
  {\cal L}_{\text{Yukawa}} =& - \overline{Q_L}
 \left(y^Q_{\bf 1} \Phi_{\bf 1}+\widetilde{y}^Q_{\bf 1}
  \widetilde{\Phi}_{\bf 1}\right)Q_R
 - \frac{1}{\sqrt{24}}~\overline{Q_L}
 \left(y_{\bf 15}^Q \Phi_{\bf 15}+\widetilde{y}^Q_{\bf 15}
  \widetilde{\Phi}_{\bf 15}\right)Q_R
\nonumber \\
& - \overline{L_L}
 \left(y^L_{\bf 1} \Phi_{\bf 1}+\widetilde{y}^L_{\bf 1}
  \widetilde{\Phi}_{\bf 1}\right)L_R
 +\sqrt{\frac{3}{8}} ~\overline{L_L}
 \left(y_{\bf 15}^L \Phi_{\bf 15}+\widetilde{y}^L_{\bf 15}
  \widetilde{\Phi}_{\bf 15}\right)L_R \nonumber \\
& -y_\Delta \overline{L_R^c} \Delta L_R
 +\text{h.c.}
 ~,
 \label{eq:lagyukawa}
 \end{align}
where $c$ denotes the charge conjugation and $\widetilde{\Phi}_{\bf 1,
15}\equiv\tau_2\Phi^*_{\bf 1, 15}\tau_2$. Here, we note that $y_{\bf 1,
15}^Q = y_{\bf 1, 15}^L$ and $\widetilde{y}_{\bf 1,
15}^Q = \widetilde{y}_{\bf 1, 15}^L$ hold at the first intermediate
scale because of the SU(4)$_C$ symmetry.
After the bi-doublet fields acquire VEVs, $ \langle \Phi_{{\bf
1, 15}}\rangle  = \text{diag} (v_u^{{\bf 1, 15}}, v_d^{{\bf 1, 15}})$,
the operators including these VEVs generate a Dirac mass
term for the charged SM fields, while the VEV of $\Delta$ gives rise to a Majorana mass terms for the
right-handed neutrinos. Since both the $\Phi_{\bf 1, 15}$ and $\Phi_{\bf
1, 15}^*$ can couple to the SM fermions, with different Yukawa couplings,
the form of the Yukawa couplings is similar to that of a generic
two-Higgs doublet model\footnote{Although there are 4 Higgs doublets below the TeV scale contributing to electroweak symmetry breaking, the SU(2)$_R$ symmetry reduces the number of Yukawa couplings.}. This Yukawa structure in general suffers from
large flavor-changing neutral currents \cite{Bjorken:1977vt,
McWilliams:1980kj, Shanker:1981mj}\footnote{However, such a generic flavor structure may explain
a 2.4$\sigma$ excess in the $h\to \mu \tau$ decay process recently
reported by the CMS collaboration \cite{Khachatryan:2015kon}. For
interpretations of this excess based on the generic two-Higgs doublet
model, see Ref.~\cite{Sierra:2014nqa}. }. In this work, we just assume
that the Yukawa couplings are appropriately
aligned so that the flavor-changing processes induced by the exchange of
the Higgs fields are  sufficiently suppressed.

As mentioned above, the lepton Yukawa couplings are equal to the quark
Yukawa couplings at the first intermediate scale. As a result, the Dirac
Yukawa couplings for neutrinos are sizable. This is potentially
problematic since the Majorana masses for right-handed neutrinos in our
model are at the TeV scale. To discuss this point, let us investigate
the neutrino mass terms. The masses of light neutrinos in our model are
given by the ordinary seesaw relation \cite{Minkowski:1977sc},
\begin{equation}
m_{\nu}\simeq - m_D
m_{\nu_R}^{-1} m_D^T~,
\end{equation}
with
\begin{align}
 m_{\nu_R} &= y_\Delta v_R^{\bf 10} ~, ~~~~~~
 m_D = (y_{\bf 1}^L v_u^{\bf 1}+ \widetilde{y}^L_{\bf 1}v_d^{\bf 1})
-\sqrt{\frac{3}{8}}(y_{\bf 15}^L v_u^{\bf 15}+ \widetilde{y}^L_{\bf 15}v_d^{\bf
 15}) ~.\label{eq:neumass}
\end{align}
The first expression in Eq.~\eqref{eq:neumass} induces Majorana masses of
${\cal O}(v_R^{\bf 10})$. Then, if the lepton Yukawa couplings are as
large as the quark Yukawa couplings, $m_D$ becomes as large as quark
masses and thus light neutrino masses generically
become much larger than ${\cal O}(1)$~eV. To get small neutrino masses,
in this paper, we suppress $m_D$ by fine-tuning the lepton Yukawa couplings.
Notice that this fine-tuning is possible because the VEV of the $({\bf
15}, {\bf 2}, {\bf 2})$ component breaks both the SU(4)$_C$ and
SU(2)$_R$ relations \cite{Babu:1992ia}. The breaking of the SU(4)$_C$ relation
can be seen in the different factors in front of $y_{\bf 15}^Q$ and
$y_{\bf 15}^L$, while the breaking of the SU(2)$_R$ relation is realized
when $v_u^{\bf 15} \neq v_d^{\bf 15}$. This allows us
to choose the neutrino Dirac Yukawa couplings freely. In this
sense, the $({\bf 15}, {\bf 2}, {\bf 2})$ component is a necessary
ingredient to make this model viable\footnote{The same trick can be used to resolve
discrepancies among the charged lepton and quark mass relations.}.

As we discuss below, we take right-handed neutrino masses to be larger
than $m_{W_R}$ to forbid the leptonic decay channels of $W_R^\pm$. On
the other hand, the right-handed neutrino masses are induced by the
Yukawa coupling $y_\Delta$.  To
satisfy this condition with a perturbative $y_\Delta$, we have a lower
bound on $v_R^{\bf 10}$, which then implies a lower bound on the $Z_R$
mass, as can be seen from Eq.~\eqref{eqn:MZR}.  For instance, for $m_{W_R} =1.9$~TeV and $y_\Delta <1$, we
have $m_{Z_R}>2.2$~TeV. The gray shaded region in Fig.~\ref{fig:zmass2}
shows the area that results in small right-handed neutrino masses. As can be seen from this figure, the LHC constraints on $M_{Z_R}$ are more severe.

Before concluding this section, we briefly discuss constraints on $W_R$
from flavor observables. The presence of $W_R$ in general induces
additional flavor violation. Currently, the $K_L$--$K_S$ mass difference
gives the most severe constraint on the mass of $W_R$, which is roughly
given by \cite{Zhang:2007da}
\begin{equation}
 m_{W_R} \gtrsim \left(\frac{g_{2R}}{g_{2L}}\right) \times
  2.5~\text{TeV}
\simeq 1.7~\text{TeV} ~,
\end{equation}
where we have set $g_{2R}/g_{2L} \simeq 0.67$, which is predicted in
our model (see Eq.~\eqref{eq:gaugcoup3221}). Hence, a $\sim 2$~TeV
$W_R$ is still allowed by flavor bounds.

%%%%%%%%%%%%%%%%%%%%%%%%%%%%%%%%%%%%%%%
\section{Diboson signal}
\label{sec:dibosonsignal}
%%%%%%%%%%%%%%%%%%%%%%%%%%%%%%%%%%%%%%

The ATLAS collaboration has reported an anomalous excess of events in the
search for resonances decaying into a pair of gauge bosons, focussing on the
hadronically decaying gauge bosons \cite{Aad:2015owa}. These gauge
bosons are detected as a fat jet since they come from a
heavy resonance and are highly boosted and thus quarks from the gauge boson
are reconstructed as a single jet with a large jet radius. By investigating invariant mass
distributions constructed from two fat jets, the ATLAS collaboration
has observed a narrow resonance around 2~TeV with $WZ$,
$WW$, and $ZZ$ final states having a local significance of 3.4, 2.6, and
2.9$\sigma$, respectively.\footnote{We note that possible subtleties in
this ATLAS analysis have been pointed out in
Ref.~\cite{Goncalves:2015yua}. For a subsequent study on this fat-jet
analysis performed by the ATLAS collaboration, see
Ref.~\cite{Aad:2015rpa}. } The tagging selections for each channel are
inexact so that sizable events are expected to contaminate other
channels. The CMS collaboration also performed a
similar search with no discrimination between $W$ and $Z$ bosons, and
found a small excess around 1.9~TeV
\cite{Khachatryan:2014hpa}. Recently, the ATLAS collaboration provided an
analysis \cite{ATLAS-CONF-2015-045} that combines searches for diboson
resonances decaying into fully leptonic \cite{Aad:2014pha},
semi-leptonic \cite{Aad:2014xka, Aad:2015ufa}, and hadronic final states
\cite{Aad:2015owa}. This still exhibits a 2.5$\sigma$ deviation around
2~TeV in the $WZ$ channel.
%{\bf Is this dominated by the hadronic modes?}.
The CMS experiment has also presented their semi-leptonic
search result and found a slight excess around 1.8~TeV
\cite{Khachatryan:2014gha}, although the collaboration found no
deviation for the leptonic decay channel\cite{Khachatryan:2014xja}. In light of these results, we here
discuss the possibility that $W_R^\pm$ in our model accounts for the
excesses observed in the above searches.

The $W_R^\pm$ can decay into $W^\pm$ and $Z$ through mixing with
$W^\pm$, and thus can yield diboson signals \cite{Hisano:2015gna, Cheung:2015nha, Gao:2015irw,
Brehmer:2015cia, Dev:2015pga, Deppisch:2015cua}. In addition, it can decay
into a pair of quarks as well as $W^\pm h$. If right-handed neutrinos
are lighter than the $W_R^\pm$ mass, the $\ell_R^\pm \nu_R$ channels are also
open, otherwise the $W_R^\pm$ does not have leptonic decay channels. The
dijet, third-generation-quark, and $W^\pm h$ resonance searches are also
important for determining if $W_R^\pm$ can explain the diboson
anomaly. As for the dijet resonance searches, the CMS experiment
\cite{Khachatryan:2015sja} observed a $\sim 2\sigma$ excess around
1.8~TeV, while the ATLAS collaboration found no significant excess
around this region \cite{Aad:2014aqa}. The CMS collaboration also
reported a 2.2$\sigma$ excess around 1.8~TeV in the $W^\pm h$ channel,
where the $W^\pm$ decays leptonically and the Higgs boson $h$ decays to
$b\overline{b}$ \cite{CMS:2015gla}. However, the ATLAS collaboration did
not observe any deviation from the background \cite{Aad:2015yza}. The CMS experiment also analyzed
the $W^\pm h$ final states in the fully hadronic channel
\cite{Khachatryan:2015bma} and found no significant excess.
Finally, the searches
in the top and bottom quark final states exhibit no excess around
$\sim 2$~TeV \cite{Aad:2014xra, Aad:2014xea, Chatrchyan:2014koa,
Khachatryan:2015edz}.

Considering these results, the authors in Ref.~\cite{Brehmer:2015cia}
carried out a cross-section fit to the data for a resonance in the
vicinity of 1.8--1.9~TeV; the resultant cross sections they obtained
are\footnote{Here, we show the result for $\sigma (pp\to W_R \to tb)$
without using the ATLAS $\ell \nu b\overline{b}$ result
\cite{Aad:2014xea}. As discussed in Ref.~\cite{Brehmer:2015cia}, this
ATLAS search systematically gives limits on the $W_R$ cross sections which are about
$2\sigma$ smaller than the expected limit for the entire mass range consider. This leads to a strong upper limit on $\sigma (pp\to W_R
\to tb)$ after the fit of 38~fb which is due to the fitting method used in that
work.  Considering the possibility of a systematic error in the ATLAS $\ell \nu
b\overline{b}$ search with its large effect on
the cross section fit shown in Ref.~\cite{Brehmer:2015cia}, we use the cross
sections found when the ATLAS $\ell \nu b\overline{b}$ search is ignored. }
\begin{align}
 \sigma (pp\to W_R \to WZ) &= 5.7^{+3.6}_{-3.3} ~\text{fb} ~,\\
 \sigma (pp\to W_R \to Wh) &= 4.5^{+5.2}_{-4.0} ~\text{fb} ~,\\
 \sigma (pp\to W_R \to jj) &= 91^{+53}_{-45} ~\text{fb} ~,\\
 \sigma (pp\to W_R \to tb) &= 0^{+39}_{-0} ~\text{fb} ~.
\end{align}

In our model, the production cross section of $W^\pm_R$ is given as a
function of the $W^\pm_R$ mass, $m_{W_R}$, and the SU(2)$_R$ gauge coupling,
$g_{2R}$. The former is determined by the position of the resonance, while
the latter is determined by enforcing gauge coupling unification and is given in Eqs.~\eqref{eq:gaugcoup3221}. The partial decay widths of the dijet and
$tb$ channels are also determined by these two quantities. On the other
hand, the partial decay widths of the $WZ$ and $Wh$ channels depend on
the $W$--$W_R$ mixing angle $\phi^W_{{LR}}$ defined by,
\begin{equation}
 \begin{pmatrix}
  W_L \\ W_R
 \end{pmatrix}
=
\begin{pmatrix}
 \cos \phi^W_{{LR}} & - \sin \phi^W_{{LR}} \\
 \sin \phi^W_{{LR}} & \cos \phi^W_{{LR}}
\end{pmatrix}
\begin{pmatrix}
 W \\ W^\prime
\end{pmatrix}
~.
\end{equation}
The size of the mixing angle is expected to be $\phi^W_{{LR}}
={\cal O}(m_W^2/m_{W_R}^2) \sim 10^{-3}$, though it is highly dependent on
the Higgs sector of the model. In the minimal left-right symmetric
model\cite{Pati:1974yy}, which is the same as $v_R^{\bf 15}=0$ in our model, the mixing angle is given by
\begin{equation}
 \phi^W_{{LR}} \simeq \sin 2\beta \left(\frac{g_{2R}}{g_{2L}}\right)
\frac{m_W^2}{m_{W_R}^2} ~,
\end{equation}
where $\tan \beta$ is the ratio of the VEVs of the bi-doublet Higgs
field $({\bf 1}, {\bf 2}, {\bf 2})_C$. To explain the observed top and
bottom quark mass ratio, $\tan \beta \simeq m_t/m_b$ is usually favored
in this model.  In our model, on the other hand, both $\Phi_{\bf 1}$
and $\Phi_{\bf 15}$ contribute to the mixing angle with a similar form
to the above expression. Contrary to the minimal left-right symmetric case, the
VEVs of these fields can be chosen almost arbitrary thanks to the
generic structure of the Yukawa sector; we can always reproduce the quark and
lepton masses by appropriately choosing these VEVs and Yukawa
couplings. Because of this added freedom in our model, we do not
specify the Higgs and Yukawa sectors of our models and regard the mixing
angle as just a free parameter of order $10^{-3}$. To simplify our discussion, we
also assume the right-handed neutrino masses are larger than the $W_R$ mass in
order to forbid the $\ell_R^\pm \nu_R$ decay channels. At the end of
this section, we briefly discuss the case in which a right-handed neutrino
has a mass lighter than $m_{W_R}$.

The authors in Ref.~\cite{Brehmer:2015cia} performed a parameter fitting
in an $\text{SU}(2)_L \otimes \text{SU}(2)_R \otimes \text{U}(1)_{B-L}$
model to determine the model parameters,
$g_{2R}/g_{2L}$ and $\phi^W_{{LR}}$, for which the required cross
sections given above are satisfied. Since the $W_R$ sector in our model
is the same as that in Ref.~\cite{Brehmer:2015cia}, we can directly
apply their result to our case. According to their 68\% CL result,
\begin{align}
 0.5\lesssim g_{2R}/g_{2L}\lesssim 0.65~,~~~~~~
 0.0012\lesssim \sin \phi^W_{{LR}} \lesssim 0.0016 ~,
\label{eq:fitparam}
\end{align}
where they took $m_{W_R} =1.9$~TeV, the value preferred by the current experimental
data. Notice that our model predicts $g_{2R}/g_{2L} \simeq 0.67$ as can be seen in Eqs.~\eqref{eq:gaugcoup3221}, and is very close to the
preferred region. Therefore, by choosing the mixing angle in the range
presented in Eq.~\eqref{eq:fitparam}, our model can explain the ATLAS diboson anomaly without conflicting with other collider bounds.

%%%%%%%%%%%%%%%%%%%%%%%%%%%%%%%%%%%%%%%%%%%%%%%%%%%%%%%%%%%%%%%%

Using the gauge couplings given in Eq.~\eqref{eq:gaugcoup3221}, and a
mixing angle in the range found in Eq.~\eqref{eq:fitparam}, we evaluate the production
cross section $\sigma(pp\to W_R^\pm)$, total decay width
$\Gamma_{\text{tot}}$, and the branching ratios of $W^\pm_R$ which
are given in the Appendix. The
total decay width and branching fractions for our model are
summarized in Table.~\ref{tab:wrsummary2}. Here, we set $m_{W_R} = 1.9$~TeV. Notice that
$\text{BR}(WZ)$ is sizable even though the mixing
angle $\sin \phi^W_{{LR}}$ is very small. This is because the
$W^\pm_R \to W^\pm Z$ decay process is enhanced by a factor of
$(m_{W_R}/m_W)^4$ due to the high-energy behavior of the longitudinal
mode of $W^\pm_R$. This enhancement compensates for the suppression coming from the mixing
angle. In addition, we have $\text{BR}(WZ)\simeq \text{BR}(Wh)$, as a
consequence of the equivalence theorem.
As for the production cross section, we compute
it using {\sc MadGraph5}
\cite{Alwall:2014hca} at the leading order, and re-scale by the so-called
$k$ factor which corresponds to the effects of the higher-order corrections;
this value is found to be $k
\simeq 1.3$ \cite{Cao:2012ng, Jezo:2014wra}. The resultant cross section
is
\begin{equation}
 \sigma (p p \to W^\pm_R) \simeq 143\times
  \left(\frac{g_{2R}}{0.42}\right)^2 ~\text{fb}  ~,
\end{equation}
again for $m_{W_R} = 1.9$~TeV. Considering the fact
that about a half of $WZ$ events decay hadronically, and using the acceptance
$\sim 0.14$ \cite{Aad:2015owa}, we estimate that there are about 9 (5) additional events in total in the presence
of $W_R^\pm$ if $\sin \phi^W_{{LR}} = 0.0016$ (0.0012). This can explain
the diboson excess observed in Ref.~\cite{Aad:2015owa}.

%%%%%%%%%%%%%%%%%%%%%%%% TABLE %%%%%%%%%%%%%%%%%%%%%%%%%%%%%%%%

\begin{table}[h]
\caption{Total decay
 width $\Gamma_{\text{tot}}$ and the branching ratios of
 $W^\pm_R$ for our model. Here, we set $m_{W_R} = 1.9$~TeV and use the
 gauge couplings given in Eq.~\eqref{eq:gaugcoup3221}.
}
\label{tab:wrsummary2}
\begin{center}
\begin{tabular}{c|c|cccc}
 \hline
 \hline
 $\sin \phi^W_{{LR}}$ & $\Gamma_{\text{tot}}$ [GeV] & $\text{BR}(jj)$ &
 $\text{BR}(tb)$ & $\text{BR}(WZ)$ & $\text{BR}(Wh)$ \\
\hline
 0.0016& 22.3 &0.61 &0.30 &0.047 &0.046 \\
% 0.08977 pb at L.O.
 0.0012& 21.4& 0.63& 0.31&0.027 &0.027 \\
 \hline
 \hline
\end{tabular}
\end{center}
\end{table}

%%%%%%%%%%%%%%%%%%%%%%%%%%%%%%%%%%%%%%%%%%%%%%%%%%%%%%%%%%%%%%%%

As we have seen above, to explain the diboson anomaly, the mixing angle
$\phi^W_{{LR}}$ should be ${\cal O}(10^{-3})$. This size of the $W$--$W_R$
mixing could be constrained by the electroweak precision measurements
\cite{Hisano:2015gna, Gao:2015irw}. Indeed, the authors in
Ref.~\cite{Grojean:2011vu} derived an upper limit on the mixing angle
from the bound on the $T$-parameter of $|\phi^W_{{LR}}| \lesssim 5\times
10^{-4}$ for $m_{W_R} = 2$~TeV, which is smaller than the values in
Eq.~\eqref{eq:fitparam}. However, our model also contains $Z_R$, which
affects the electroweak observables as well. Moreover, since its
contribution modifies the $Z$-boson coupling to the SM fermions at tree
level through the $Z$--$Z_R$ mixing, we cannot use the conventional
method relying on the $S$ and $T$ parameters \cite{Peskin:1990zt} to evaluate the electroweak
precision constraints. Instead, we need to perform a complete parameter fitting
onto the electroweak observables. This is beyond the scope of this
paper. We however note that such a parameter fitting has been carried
out for the $\text{SU}(2)_L \otimes \text{SU}(2)_R \otimes
\text{U}(1)_{B-L}$ models in the literature \cite{Cao:2012ng,
Hsieh:2010zr} and, according to the results, a 2~TeV $W_R$ with an
${\cal O}(10^{-3})$ $W$--$W_R$ mixing can evade the electroweak
precision constraint. Since the structure of the TeV-scale gauge sector
of our model is similar to those in these models, we expect that our
model can also avoid this constraint.

Next, we consider the case where a
right-handed neutrino has a mass lighter than the $W_R$ mass. This case
is potentially interesting since it may explain the 2.8$\sigma$ anomaly
observed by the CMS collaboration in the right-handed neutrino searches looking for dijet plus dilepton events \cite{Khachatryan:2014dka}. Indeed, it
turns out that $W_R$ with $g_{2R}/g_{2L} \simeq 0.6$ and $m_{W_R} \sim
2$~TeV may explain the excess if the $W_R$ decays into a
right-handed neutrino that mainly couples to first generation charged leptons
\cite{Deppisch:2014qpa}, i.e. $W_R^\pm\to \ell^\pm N_R\to \ell^\pm \ell ^\pm jj$. This simple setup is, however, disfavored
once the charge of observed electrons/positrons is considered. The CMS
collaboration observed only one same-sign electron event on top of the 13
opposite-sign events. However, if the right-handed neutrino is
a Majorana fermion, we expect almost the same numbers of events for each
case\footnote{Since a Majorana fermion and its anti-particle are identical, if a
Majorana fermion decays into an electron it would also decay into a
positron with the same rate.}. In addition, such a Majorana right-handed neutrino is severely
restricted by the ATLAS search for same-sign leptons plus dijet events
\cite{Aad:2015xaa}. To reconcile these experimental results, there are several
possibilities for extending the neutrino sector that work, such as the
inverse seesaw \cite{Mohapatra:1986aw} or the linear seesaw
\cite{Barr:2003nn} mechanisms which were discussed in the context of the diboson anomaly in Refs.~\cite{Dev:2015pga} and
\cite{Deppisch:2015cua}, respectively. In these cases, heavy neutrinos
become pseudo-Dirac fermions, which allows them to evade the bounds from
the same-sign leptons plus dijet searches \cite{Khachatryan:2014dka,
Aad:2015xaa}. Although our model potentially accommodates such an
extension, we leave the investigation of this to future work.

Finally, we discuss future prospects for testing this model. The first
step for verifying our model is, of course, to confirm the presence of
a $\sim 2$~TeV $W_R$ at Run-II of the LHC . The production cross section of
a $W_R$ in our model at a $13$~TeV center-of-mass energy is estimated as
$\sim 800$~fb, where we have used the $k$ factor of $k \simeq 1.2$
\cite{Cao:2012ng, Jezo:2014wra}. Therefore, a $\sim 2$~TeV $W_R$ can be
tested with an integrated luminosity of $\sim 10$~fb$^{-1}$. As
discussed above, our model may predict a relatively light $Z_R$, which
may also be reached at the LHC. In addition, a (fat) jet plus missing
energy search can also probe this model \cite{Liew:2015osa}, which would constrain the decay mode $WZ\to \nu\bar \nu jj$.

The presence of an SU(2)$_L$ triplet Dirac fermion dark matter is a
distinct feature of our model, and thus dark matter searches can also
play an important role in testing this model. Since we expect its mass
to be around 2~TeV, it is difficult for the LHC to probe it
directly. Instead, indirect dark matter detection experiments, such as
searches for gamma rays from the Galactic Center or dwarf spheroidal
galaxies, are quite promising since this dark matter has a relatively
large annihilation cross section. Most work to date on an SU(2)$_L$ triplet
fermion dark matter candidate has assumed a Majorana rather than a Dirac particle \cite{Cohen:2013ama} and there is little work on the Dirac triplet
dark matter. To test the dark matter in our model in the indirect
detection experiments, it is quite important to evaluate its
annihilation cross section precisely and compare it with that of the
Majorana dark matter. The direct detection rate of the SU(2)$_L$ triplet
dark matter is rather small; its scattering cross section with a
nucleon is computed as $\sigma_{\text{SI}} \simeq 2\times 10^{-47}$~cm$^{2}$
\cite{Hisano:2015rsa}. Still, it is in principle detectable in future
experiments, which gives another way to test our model.

%%%%%%%%%%%%%%%%%%%%%%%%%%%%%%%%%%%%%%%%%%
\section{Conclusion}
\label{sec:conclusion}
%%%%%%%%%%%%%%%%%%%%%%%%%%%%%%%%%%%%%%%%%%

We have considered a specific (non-supersymmetric) SO(10) model
which can simultaneously explain the purported ATLAS diboson excess,
provide a stable dark matter candidate, and achieve gauge coupling unification
with a suitably high GUT scale. In addition to the gauge and matter sector of the
theory which are standard for SO(10), the model has Higgs fields with three different representations under the SO(10) gauge symmetry: a {\bf 210} which is used to break SO(10), and plays an role in breaking both of the intermediate scale
gauge groups, $\text{SU}(4)_C \otimes \text{SU}(2)_L \otimes
\text{SU}(2)_R$ and $\text{SU}(3)_C \otimes \text{SU}(2)_L \otimes
\text{SU}(2)_R \otimes \text{U}(1)_{B-L}$; a {\bf 126} which assist the breaking of
the smaller intermediate gauge group and participates in the breaking of the SM gauge symmetries; a {\bf 10} which is also involved in the breaking of the SM gauge symmetries. The model also introduces a Dirac {\bf 45} which
provides us with a stable dark matter candidate in the form of a $ \text{SU}(2)_L$ triplet.
Having fixed the lower intermediate scale at 1.9 TeV (to account for the diboson excess),
the GUT scale of order $10^{17}$ GeV and the high intermediate scale of order $10^9$ GeV
are determined by requiring gauge coupling unification. In this model, both stages of
intermediate symmetry breaking as well as a TeV-scale dark matter triplet are necessary
for unification of the gauge couplings.

The high GUT scale ensures proton stability. The relatively high intermediate scale of
$10^9$ GeV, ensures that the effects induced by the presence of both vector and scalar leptoquarks
are small enough to remain compatible with experimental constraints. The lower
intermediate scale was chosen to account for the diboson excess, and predicts
new right-handed gauge bosons to appear at the TeV scale.
In addition to a charged $W_R^\pm$ pair, the model predicts a new $Z_R$ gauge boson
whose mass depends on the relative contributions of the two Higgs VEVs participating in the breaking of the low intermediate
scale. This set of TeV scale gauge bosons should allow the model to be
tested in current and future runs at the LHC.

The contributions of two Higgs VEVs to SM breaking
gives the model the flexibility needed to adjust the SM fermion spectrum, thus avoiding
some of the overly restrictive predictions often encountered in minimal SO(10) models.
This is necessary if one hopes to derive eV scale neutrino masses from the
seesaw mechanism given that the Majorana mass for the right-handed neutrino is also
at the TeV scale.

Finally, the model contains a testable dark matter candidate and the neutral member of an electroweak triplet.
Similar to the often studied supersymmetric wino, the mass splitting is rather small (typically of
order 165 MeV), and the annihilation cross section is rather large, requiring the mass of the triplet
to be of order 2 TeV as well.  Thus the dark matter candidate in this model is testable in indirect dark matter searches as well as direct detection experiments.

~\\~\\
{\it Note Added:}
While completing this work, the CMS collaboration announced a new result
for the dijet resonance search using the 13~TeV LHC data with an
integrated luminosity of 2.4~fb$^{-1}$ \cite{Khachatryan:2015dcf}. They
give an upper bound on the product of the production cross section
($\sigma$), the branching fraction of the dijet channel
($\text{Br}(jj)$), and acceptance ($A$) of $\sigma A \text{Br}(jj)
\lesssim 400$~fb for a resonance mass of $1.9$~TeV, and thus we have
$\sigma \text{Br}(jj) \lesssim 667$~fb using $A\simeq 0.6$
\cite{Khachatryan:2015dcf}. The ATLAS collaboration also reported a 13-TeV
result for the same channel based on the 3.6~fb$^{-1}$ data, and gave a
more severe constraint \cite{ATLAS:2015nsi}: $\sigma A\text{Br}(jj)
\lesssim 210$~fb for a 1.9~TeV resonance. By using $A\simeq 0.4$
\cite{ATLAS:2015nsi}, we obtain $\sigma \text{Br}(jj) \lesssim 525$~fb.
On the other hand, our model predicts $\sigma
\text{Br}(jj)$ to be $\sigma \text{Br}(jj) = 421$ (485)~fb for a
$1.9$~TeV $W_R$ with $\sin\phi^W_{LR} = 0.0016$ ($0.0012$),
and thus evades these limits. Since these limits are fairly close to the model
predictions, we expect that the presence of $W_R$ in our model can be
confirmed in near future.

%%%%%%%%%%%%%%%%%%%%%%%%%%%%%%%%%%%%
\section*{Acknowledgments}
%%%%%%%%%%%%%%%%%%%%%%%%%%%%%%%%%%%%

The work of J.L.E., N.N. and K.A.O. was supported by DOE grant
DE-SC0011842 at the University of Minnesota.

%%%%%%%%%%%%%%%%%%%%%%%%%%%%%%%%%%%%%%%%%%%%%%
\section*{Appendix}
\appendix
%%%%%%%%%%%%%%%%%%%%%%%%%%%%%%%%%%%%%%%%%%%%%

%%%%%%%%%%%%%%%%%%%%%%%%%%%%%%%%%%%%%%%
\section{Decay widths}
\label{sec:decaywidths}
%%%%%%%%%%%%%%%%%%%%%%%%%%%%%%%%%%%%%%%

Here, we summarize the partial decay widths of $W_R$ we use in our
calculation. For the fermionic channels, $W_R \to f \bar{f}^\prime$, we
use
\begin{align}
 \Gamma (W_R^{+}\to u \bar{d})
&= \Gamma (W_R^{+}\to c \bar{s})
=\frac{g_{2R}^2}{16\pi}
m_{W_R}~, \\
 \Gamma (W_R^{+}\to t \bar{b})
&=\frac{g_{2R}^2}{16\pi}
m_{W_R} \left(1+\frac{m_t^2}{2 m_{W_R}^2}\right)
\left(1-\frac{m_t^2}{m_{W_R}^2}\right)^2~.
\end{align}
For the $W_R\to WZ$ decay process, we have
\begin{align}
 \Gamma(W_R^+ \to W^+ Z)
=& \frac{g_{2L}^2}{192\pi}\sin^2 \phi^W_{LR} \frac{m_{W_R}^5}{m_W^4}
\left(1-2
 \frac{m_W^2+m_Z^2}{m_{W_R}^2}+\frac{(m_W^2-m_Z^2)^2}{m_{W_R}^4}
\right)^{\frac{3}{2}}  \nonumber \\
&\times \left(
1+10 \frac{m_W^2+m_Z^2}{m_{W_R}^2} +
\frac{m_W^4 +10 m_W^2 m_Z^2 + m_Z^4}{m^4_{W_R}}
\right)~.
\end{align}
Finally, the $W_R\to Wh$ decay width is given by
\begin{align}
 \Gamma(W_R^+ \to W^+ h)
=& \frac{g_{2L}^2}{192\pi}\sin^2 \phi^W_{LR} \frac{m_{W_R}^5}{m_W^4}
\left(1-2
 \frac{m_W^2+m_h^2}{m_{W_R}^2}+\frac{(m_W^2-m_h^2)^2}{m_{W_R}^4}
\right)^{\frac{1}{2}}  \nonumber \\
&\times \left(
1+\frac{10m_W^2-2m_h^2}{m_{W_R}^2} +
\frac{(m_W^2-m_h^2)^2}{m^4_{W_R}}
\right)~,
\end{align}
in the decoupling limit.

%%%%%%%%%%%%%%%%%%%%%%%%%%%%%%%%%%%%%%%

%%%%%%%%%%%%%%%%%%%%%%%%%%%%%%%%%%%%%%%


\begin{thebibliography}{99}

%\cite{Georgi:1974sy}
\bibitem{Georgi:1974sy}
  H.~Georgi and S.~L.~Glashow,
  %``Unity of All Elementary Particle Forces,''
  Phys.\ Rev.\ Lett.\  {\bf 32}, 438 (1974).
  %%CITATION = PRLTA,32,438;%%
  %3981 citations counted in INSPIRE as of 29 Aug 2015

%\cite{Georgi:1974my}
\bibitem{Georgi:1974my}
  H.~Georgi,
  %``The State of the Art—Gauge Theories,''
  AIP Conf.\ Proc.\  {\bf 23}, 575 (1975).
  %%CITATION = APCPC,23,575;%%
  %116 citations counted in INSPIRE as of 29 Aug 2015
%\cite{Fritzsch:1974nn}
%\bibitem{Fritzsch:1974nn}
  H.~Fritzsch and P.~Minkowski,
  %``Unified Interactions of Leptons and Hadrons,''
  Annals Phys.\  {\bf 93}, 193 (1975).
  %%CITATION = APNYA,93,193;%%
  %1395 citations counted in INSPIRE as of 29 Aug 2015

  \bibitem{so10-2}
   M.~S.~Chanowitz, J.~R.~Ellis and M.~K.~Gaillard,
  %``The Price of Natural Flavor Conservation in Neutral Weak Interactions,''
  Nucl.\ Phys.\ B {\bf 128}, 506 (1977);
  %%CITATION = NUPHA,B128,506;%%
   H.~Georgi and D.~V.~Nanopoulos,
  %``Ordinary Predictions from Grand Principles: T Quark Mass in O(10),''
  Nucl.\ Phys.\ B {\bf 155}, 52 (1979);
  %%CITATION = NUPHA,B155,52;%%

  \bibitem{GN2}
  H.~Georgi and D.~V.~Nanopoulos,
  %``Masses and Mixing in Unified Theories,''
  Nucl.\ Phys.\ B {\bf 159}, 16 (1979);
  %%CITATION = NUPHA,B159,16;%%
  C.~E.~Vayonakis,
  %``On Mass Relations and Renormalization Effects in Grand Unified Theories,''
  Phys.\ Lett.\ B {\bf 82}, 224 (1979)
  [Phys.\ Lett.\  {\bf 83B}, 421 (1979)].
  %%CITATION = PHLTA,B82,224;%%


%\cite{Rajpoot:1980xy}
\bibitem{Rajpoot:1980xy}
  S.~Rajpoot,
  %``Symmetry Breaking And Intermediate Mass Scales In The So(10) Grand Unified Theory,''
  Phys.\ Rev.\ D {\bf 22}, 2244 (1980);
  %%CITATION = PHRVA,D22,2244;%%
  %73 citations counted in INSPIRE as of 05 Feb 2015
%
%\cite{Yasue:1981nd}
%\bibitem{Yasue:1981nd}
  M.~Yasue,
  %``Phenomenological Aspect of SO(10) Grand Unified Model,''
  Prog.\ Theor.\ Phys.\  {\bf 65}, 708 (1981)
  [Erratum-ibid.\  {\bf 65}, 1480 (1981)];
  %%CITATION = PTPKA,65,708;%%
  %7 citations counted in INSPIRE as of 05 Feb 2015
%\cite{Gipson:1984aj}
%\bibitem{Gipson:1984aj}
  J.~M.~Gipson and R.~E.~Marshak,
  %``Intermediate Mass Scales in the New SO(10) Grand Unification in the One Loop Approximation,''
  Phys.\ Rev.\ D {\bf 31}, 1705 (1985);
  %%CITATION = PHRVA,D31,1705;%%
  %39 citations counted in INSPIRE as of 05 Feb 2015
%\cite{Chang:1984qr}
%\bibitem{Chang:1984qr}
  D.~Chang, R.~N.~Mohapatra, J.~Gipson, R.~E.~Marshak and M.~K.~Parida,
  %``Experimental Tests of New SO(10) Grand Unification,''
  Phys.\ Rev.\ D {\bf 31}, 1718 (1985);
  %%CITATION = PHRVA,D31,1718;%%
  %157 citations counted in INSPIRE as of 06 Feb 2015
%\cite{Deshpande:1992au}
%\bibitem{Deshpande:1992au}
  N.~G.~Deshpande, E.~Keith and P.~B.~Pal,
  %``Implications of LEP results for SO(10) grand unification,''
  Phys.\ Rev.\ D {\bf 46}, 2261 (1993);
  %%CITATION = PHRVA,D46,2261;%%
  %83 citations counted in INSPIRE as of 06 Feb 2015
%\cite{Deshpande:1992em}
%\bibitem{Deshpande:1992em}
  N.~G.~Deshpande, E.~Keith and P.~B.~Pal,
  %``Implications of LEP results for SO(10) grand unification with two intermediate stages,''
  Phys.\ Rev.\ D {\bf 47}, 2892 (1993)
  [hep-ph/9211232];
  %%CITATION = HEP-PH/9211232;%%
  %35 citations counted in INSPIRE as of 06 Feb 2015
%\cite{Bertolini:2009es}
%\bibitem{Bertolini:2009es}
  S.~Bertolini, L.~Di Luzio and M.~Malinsky,
  %``On the vacuum of the minimal nonsupersymmetric SO(10) unification,''
  Phys.\ Rev.\ D {\bf 81}, 035015 (2010)
  [arXiv:0912.1796 [hep-ph]].
  %%CITATION = ARXIV:0912.1796;%%
  %27 citations counted in INSPIRE as of 06 Feb 2015


  %\cite{Fukugita:1993fr}
\bibitem{Fukugita:1993fr}
  M.~Fukugita and T.~Yanagida,
  %``Physics of neutrinos,''
  In *Fukugita, M. (ed.), Suzuki, A. (ed.): Physics and astrophysics of neutrinos* 1-248. and Kyoto Univ. - YITP-K-1050 (93/12,rec.Feb.94) 248 p. C.
  %4 citations counted in INSPIRE as of 06 Feb 2015

%\cite{DiLuzio:2011my}
\bibitem{DiLuzio:2011my}
  L.~Di Luzio,
  %``Aspects of symmetry breaking in Grand Unified Theories,''
  arXiv:1110.3210 [hep-ph].
  %%CITATION = ARXIV:1110.3210;%%
  %9 citations counted in INSPIRE as of 31 May 2015

%\cite{Fukuyama:2012rw}
\bibitem{Fukuyama:2012rw}
  T.~Fukuyama,
  %``SO(10) GUT in Four and Five Dimensions: A Review,''
  Int.\ J.\ Mod.\ Phys.\ A {\bf 28}, 1330008 (2013)
  [arXiv:1212.3407 [hep-ph]].
  %%CITATION = ARXIV:1212.3407;%%
  %6 citations counted in INSPIRE as of 29 Aug 2015


 \bibitem{so10-3}
  A.~Masiero,
  %``On the Phenomenological Group in Unified SO(10) Model,''
  Phys.\ Lett.\ B {\bf 93}, 295 (1980);
  %%CITATION = PHLTA,B93,295;%%
%
 % \bibitem{lsw}
  G.~Lazarides, Q.~Shafi and C.~Wetterich,
  %``Proton Lifetime and Fermion Masses in an SO(10) Model,''
  Nucl.\ Phys.\ B {\bf 181}, 287 (1981);
  %%CITATION = NUPHA,B181,287;%%
%
%  \bibitem{ssw}
  Q.~Shafi, M.~Sondermann and C.~Wetterich,
  %``Fourth Color in O(10),''
  Phys.\ Lett.\ B {\bf 92}, 304 (1980);
  %%CITATION = PHLTA,B92,304;%%
%
%  \bibitem{delAguila:1980at}
  F.~del Aguila and L.~E.~Ibanez,
  %``Higgs Bosons in SO(10) and Partial Unification,''
  Nucl.\ Phys.\ B {\bf 177}, 60 (1981);
  %%CITATION = NUPHA,B177,60;%%
%\cite{Babu:2015bna}
%\bibitem{Babu:2015bna}
  K.~S.~Babu and S.~Khan,
  %``Minimal nonsupersymmetric $SO(10)$ model: Gauge coupling unification, proton decay, and fermion masses,''
  Phys.\ Rev.\ D {\bf 92}, 075018 (2015)
%  doi:10.1103/PhysRevD.92.075018
  [arXiv:1507.06712 [hep-ph]].
  %%CITATION = doi:10.1103/PhysRevD.92.075018;%%
  %3 citations counted in INSPIRE as of 02 Dec 2015

%\cite{Aad:2015owa}
\bibitem{Aad:2015owa}
  G.~Aad {\it et al.} [ATLAS Collaboration],
  %``Search for high-mass diboson resonances with boson-tagged jets in proton-proton collisions at $\sqrt{s} = 8$ TeV with the ATLAS detector,''
  arXiv:1506.00962 [hep-ex].
  %%CITATION = ARXIV:1506.00962;%%
  %53 citations counted in INSPIRE as of 29 Aug 2015

%\cite{Hisano:2015gna}
\bibitem{Hisano:2015gna}
  J.~Hisano, N.~Nagata and Y.~Omura,
  %``Interpretations of the ATLAS Diboson Resonances,''
  Phys.\ Rev.\ D {\bf 92}, no. 5, 055001 (2015)
%  doi:10.1103/PhysRevD.92.055001
  [arXiv:1506.03931 [hep-ph]].
  %%CITATION = doi:10.1103/PhysRevD.92.055001;%%
  %44 citations counted in INSPIRE as of 02 Dec 2015

%\cite{Cheung:2015nha}
\bibitem{Cheung:2015nha}
  K.~Cheung, W.~Y.~Keung, P.~Y.~Tseng and T.~C.~Yuan,
  %``Interpretations of the ATLAS Diboson Anomaly,''
  Phys.\ Lett.\ B {\bf 751}, 188 (2015)
%  doi:10.1016/j.physletb.2015.10.029
  [arXiv:1506.06064 [hep-ph]];
  %%CITATION = doi:10.1016/j.physletb.2015.10.029;%%
  %45 citations counted in INSPIRE as of 02 Dec 2015
%\cite{Dobrescu:2015qna}
%\bibitem{Dobrescu:2015qna}
  B.~A.~Dobrescu and Z.~Liu,
  %``W′ Boson near 2 TeV: Predictions for Run 2 of the LHC,''
  Phys.\ Rev.\ Lett.\  {\bf 115}, no. 21, 211802 (2015)
 % doi:10.1103/PhysRevLett.115.211802
  [arXiv:1506.06736 [hep-ph]];
  %%CITATION = doi:10.1103/PhysRevLett.115.211802;%%
  %52 citations counted in INSPIRE as of 02 Dec 2015
%\cite{Thamm:2015csa}
%\bibitem{Thamm:2015csa}
  A.~Thamm, R.~Torre and A.~Wulzer,
  %``A composite Heavy Vector Triplet in the ATLAS di-boson excess,''
  Phys.\ Rev.\ Lett.\  {\bf 115}, no. 22, 221802 (2015)
 % doi:10.1103/PhysRevLett.115.221802
  [arXiv:1506.08688 [hep-ph]];
  %%CITATION = doi:10.1103/PhysRevLett.115.221802;%%
  %39 citations counted in INSPIRE as of 02 Dec 2015
%\cite{Cao:2015lia}
%\bibitem{Cao:2015lia}
  Q.~H.~Cao, B.~Yan and D.~M.~Zhang,
  %``Simple non-Abelian extensions of the standard model gauge group and the diboson excesses at the LHC,''
  Phys.\ Rev.\ D {\bf 92}, no. 9, 095025 (2015)
%  doi:10.1103/PhysRevD.92.095025
  [arXiv:1507.00268 [hep-ph]];
  %%CITATION = doi:10.1103/PhysRevD.92.095025;%%
  %40 citations counted in INSPIRE as of 02 Dec 2015
%\cite{Abe:2015jra}
%\bibitem{Abe:2015jra}
  T.~Abe, R.~Nagai, S.~Okawa and M.~Tanabashi,
  %``Unitarity sum rules, three site moose model, and the ATLAS 2 TeV diboson anomalies,''
  Phys.\ Rev.\ D {\bf 92}, no. 5, 055016 (2015)
  [arXiv:1507.01185 [hep-ph]];
  %%CITATION = ARXIV:1507.01185;%%
  %19 citations counted in INSPIRE as of 18 sept. 2015
%\cite{Allanach:2015hba}
%\bibitem{Allanach:2015hba}
  B.~C.~Allanach, B.~Gripaios and D.~Sutherland,
  %``Anatomy of the ATLAS diboson anomaly,''
  Phys.\ Rev.\ D {\bf 92}, 055003 (2015)
  [arXiv:1507.01638 [hep-ph]];
  %%CITATION = ARXIV:1507.01638;%%
  %18 citations counted in INSPIRE as of 06 sept. 2015
%\cite{Abe:2015uaa}
%\bibitem{Abe:2015uaa}
  T.~Abe, T.~Kitahara and M.~M.~Nojiri,
  %``Prospects for Spin-1 Resonance Search at 13 TeV LHC and the ATLAS Diboson Excess,''
  arXiv:1507.01681 [hep-ph];
  %%CITATION = ARXIV:1507.01681;%%
  %20 citations counted in INSPIRE as of 29 Aug 2015
%\cite{Dobrescu:2015yba}
%\bibitem{Dobrescu:2015yba}
  B.~A.~Dobrescu and Z.~Liu,
  %``Heavy Higgs bosons and the 2 TeV W$^{′}$ boson,''
  JHEP {\bf 1510}, 118 (2015)
%  doi:10.1007/JHEP10(2015)118
  [arXiv:1507.01923 [hep-ph]];
  %%CITATION = doi:10.1007/JHEP10(2015)118;%%
  %32 citations counted in INSPIRE as of 02 Dec 2015
%\cite{Faraggi:2015iaa}
%\bibitem{Faraggi:2015iaa}
  A.~E.~Faraggi and M.~Guzzi,
  %``Extra $Z^{\prime }$ s and $W^{\prime }$ s in heterotic-string derived models,''
  Eur.\ Phys.\ J.\ C {\bf 75}, no. 11, 537 (2015)
%  doi:10.1140/epjc/s10052-015-3763-4
  [arXiv:1507.07406 [hep-ph]];
  %%CITATION = doi:10.1140/epjc/s10052-015-3763-4;%%
  %10 citations counted in INSPIRE as of 02 Dec 2015
%\cite{Coloma:2015una}
%\bibitem{Coloma:2015una}
  P.~Coloma, B.~A.~Dobrescu and J.~Lopez-Pavon,
  %``Right-Handed Neutrinos and the 2 TeV $W'$ Boson,''
  arXiv:1508.04129 [hep-ph];
  %%CITATION = ARXIV:1508.04129;%%
  %7 citations counted in INSPIRE as of 02 Dec 2015
%\cite{Collins:2015wua}
%\bibitem{Collins:2015wua}
  J.~H.~Collins and W.~H.~Ng,
  %``A 2 TeV $W_R$, Supersymmetry, and the Higgs Mass,''
  arXiv:1510.08083 [hep-ph];
  %%CITATION = ARXIV:1510.08083;%%
  %4 citations counted in INSPIRE as of 02 Dec 2015
%\cite{Dobrescu:2015jvn}
%\bibitem{Dobrescu:2015jvn}
  B.~A.~Dobrescu and P.~J.~Fox,
  %``Signals of a 2 TeV $W'$ boson and a heavier $Z'$ boson,''
  arXiv:1511.02148 [hep-ph];
  %%CITATION = ARXIV:1511.02148;%%
  %2 citations counted in INSPIRE as of 02 Dec 2015
%\cite{Sajjad:2015urz}
%\bibitem{Sajjad:2015urz}
  A.~Sajjad,
  %``Understanding diboson anomalies,''
  arXiv:1511.02244 [hep-ph];
  %%CITATION = ARXIV:1511.02244;%%
  %1 citations counted in INSPIRE as of 02 Dec 2015
%\cite{Appelquist:2015vdl}
%\bibitem{Appelquist:2015vdl}
  T.~Appelquist, Y.~Bai, J.~Ingoldby and M.~Piai,
  %``Spectrum-doubled Heavy Vector Bosons at the LHC,''
  arXiv:1511.05473 [hep-ph];
  %%CITATION = ARXIV:1511.05473;%%
%\cite{Das:2015ysz}
%\bibitem{Das:2015ysz}
  K.~Das, T.~Li, S.~Nandi and S.~K.~Rai,
  %``The Diboson Excesses in an Anomaly Free Leptophobic Left-Right Model,''
  arXiv:1512.00190 [hep-ph];
  %%CITATION = ARXIV:1512.00190;%%
%\cite{Hirsch:2015fvq}
%\bibitem{Hirsch:2015fvq}
J.~A.~Aguilar-Saavedra and F.~R.~Joaquim,
  %``Multiboson production in W' decays,''
  arXiv:1512.00396 [hep-ph];
  %%CITATION = ARXIV:1512.00396;%%
  M.~Hirsch, M.~E.~Krauss, T.~Opferkuch, W.~Porod and F.~Staub,
  %``A constrained supersymmetric left-right model,''
  arXiv:1512.00472 [hep-ph].
  %%CITATION = ARXIV:1512.00472;%%

%\cite{Gao:2015irw}
\bibitem{Gao:2015irw}
  Y.~Gao, T.~Ghosh, K.~Sinha and J.~H.~Yu,
  %``SU(2)×SU(2)×U(1) interpretations of the diboson and Wh excesses,''
  Phys.\ Rev.\ D {\bf 92}, no. 5, 055030 (2015)
 % doi:10.1103/PhysRevD.92.055030
  [arXiv:1506.07511 [hep-ph]].
  %%CITATION = doi:10.1103/PhysRevD.92.055030;%%
  %45 citations counted in INSPIRE as of 02 Dec 2015


%\cite{Brehmer:2015cia}
\bibitem{Brehmer:2015cia}
  J.~Brehmer, J.~Hewett, J.~Kopp, T.~Rizzo and J.~Tattersall,
  %``Symmetry Restored in Dibosons at the LHC?,''
  JHEP {\bf 1510}, 182 (2015)
%  doi:10.1007/JHEP10(2015)182
  [arXiv:1507.00013 [hep-ph]].
  %%CITATION = doi:10.1007/JHEP10(2015)182;%%
  %50 citations counted in INSPIRE as of 02 Dec 2015

%\cite{Dev:2015pga}
\bibitem{Dev:2015pga}
  P.~S.~Bhupal Dev and R.~N.~Mohapatra,
  %``Unified explanation of the $eejj$, diboson and dijet resonances at the LHC,''
  Phys.\ Rev.\ Lett.\  {\bf 115}, no. 18, 181803 (2015)
%  doi:10.1103/PhysRevLett.115.181803
  [arXiv:1508.02277 [hep-ph]].
  %%CITATION = doi:10.1103/PhysRevLett.115.181803;%%
  %22 citations counted in INSPIRE as of 02 Dec 2015

%\cite{Deppisch:2015cua}
\bibitem{Deppisch:2015cua}
  F.~F.~Deppisch, L.~Graf, S.~Kulkarni, S.~Patra, W.~Rodejohann, N.~Sahu and U.~Sarkar,
  %``Reconciling the 2 TeV Excesses at the LHC in a Linear Seesaw Left-Right Model,''
  arXiv:1508.05940 [hep-ph].
  %%CITATION = ARXIV:1508.05940;%%





%\cite{Khachatryan:2014dka}
\bibitem{Khachatryan:2014dka}
  V.~Khachatryan {\it et al.} [CMS Collaboration],
  %``Search for heavy neutrinos and $\mathrm {W}$ bosons with right-handed couplings in proton-proton collisions at $\sqrt{s} = 8\,\text {TeV} $,''
  Eur.\ Phys.\ J.\ C {\bf 74}, no. 11, 3149 (2014)
  [arXiv:1407.3683 [hep-ex]].
  %%CITATION = ARXIV:1407.3683;%%
  %68 citations counted in INSPIRE as of 31 Aug 2015


%\cite{Kibble:1982ae}
\bibitem{Kibble:1982ae}
  T.~W.~B.~Kibble, G.~Lazarides and Q.~Shafi,
  %``Strings in SO(10),''
  Phys.\ Lett.\ B {\bf 113}, 237 (1982).
  %%CITATION = PHLTA,B113,237;%%
  %140 citations counted in INSPIRE as of 02 Mar 2015

%\cite{Krauss:1988zc}
\bibitem{Krauss:1988zc}
  L.~M.~Krauss and F.~Wilczek,
  %``Discrete Gauge Symmetry in Continuum Theories,''
  Phys.\ Rev.\ Lett.\  {\bf 62}, 1221 (1989).
  %%CITATION = PRLTA,62,1221;%%
  %408 citations counted in INSPIRE as of 26 Jan 2015

%\cite{Ibanez:1991hv}
\bibitem{Ibanez:1991hv}
  L.~E.~Ibanez and G.~G.~Ross,
  %``Discrete gauge symmetry anomalies,''
  Phys.\ Lett.\ B {\bf 260}, 291 (1991);
  %%CITATION = PHLTA,B260,291;%%
  %261 citations counted in INSPIRE as of 26 gen 2015
%
%\cite{Ibanez:1991pr}
%\bibitem{Ibanez:1991pr}
  L.~E.~Ibanez and G.~G.~Ross,
  %``Discrete gauge symmetries and the origin of baryon and lepton number conservation in supersymmetric versions of the standard model,''
  Nucl.\ Phys.\ B {\bf 368}, 3 (1992).
  %%CITATION = NUPHA,B368,3;%%
  %469 citations counted in INSPIRE as of 26 Jan 2015

%\cite{Martin:1992mq}
\bibitem{Martin:1992mq}
  S.~P.~Martin,
  %``Some simple criteria for gauged R-parity,''
  Phys.\ Rev.\ D {\bf 46}, 2769 (1992)
  [hep-ph/9207218].
  %%CITATION = HEP-PH/9207218;%%
  %221 citations counted in INSPIRE as of 26 gen 2015


%\cite{Kadastik:2009dj}
\bibitem{Kadastik:2009dj}
  M.~Kadastik, K.~Kannike and M.~Raidal,
  %``Matter parity as the origin of scalar Dark Matter,''
  Phys.\ Rev.\ D {\bf 81}, 015002 (2010)
  [arXiv:0903.2475 [hep-ph]];
  %%CITATION = ARXIV:0903.2475;%%
  %53 citations counted in INSPIRE as of 26 gen 2015
%
%\cite{Kadastik:2009cu}
%\bibitem{Kadastik:2009cu}
  M.~Kadastik, K.~Kannike and M.~Raidal,
  %``Dark Matter as the signal of Grand Unification,''
  Phys.\ Rev.\ D {\bf 80}, 085020 (2009)
  [Erratum-ibid.\ D {\bf 81}, 029903 (2010)]
  [arXiv:0907.1894 [hep-ph]].
  %%CITATION = ARXIV:0907.1894;%%
  %41 citations counted in INSPIRE as of 26 gen 2015

%\cite{Frigerio:2009wf}
\bibitem{Frigerio:2009wf}
  M.~Frigerio and T.~Hambye,
  %``Dark matter stability and unification without supersymmetry,''
  Phys.\ Rev.\ D {\bf 81}, 075002 (2010)
  [arXiv:0912.1545 [hep-ph]];
  %%CITATION = ARXIV:0912.1545;%%
  %21 citations counted in INSPIRE as of 26 Jan 2015
%
%\cite{Hambye:2010zb}
%\bibitem{Hambye:2010zb}
  T.~Hambye,
  %``On the stability of particle dark matter,''
  PoS IDM {\bf 2010}, 098 (2011)
  [arXiv:1012.4587 [hep-ph]].
  %%CITATION = ARXIV:1012.4587;%%
  %21 citations counted in INSPIRE as of 26 gen 2015

%\cite{Mambrini:2015vna}
\bibitem{Mambrini:2015vna}
  Y.~Mambrini, N.~Nagata, K.~A.~Olive, J.~Quevillon and J.~Zheng,
  %``Dark matter and gauge coupling unification in nonsupersymmetric SO(10) grand unified models,''
  Phys.\ Rev.\ D {\bf 91}, no. 9, 095010 (2015)
  [arXiv:1502.06929 [hep-ph]].
  %%CITATION = ARXIV:1502.06929;%%
  %5 citations counted in INSPIRE as of 31 Aug 2015

%\cite{Nagata:2015dma}
\bibitem{Nagata:2015dma}
  N.~Nagata, K.~A.~Olive and J.~Zheng,
  %``Weakly-Interacting Massive Particles in Non-supersymmetric SO(10) Grand Unified Models,''
  JHEP {\bf 1510}, 193 (2015)
 % doi:10.1007/JHEP10(2015)193
  [arXiv:1509.00809 [hep-ph]].
  %%CITATION = doi:10.1007/JHEP10(2015)193;%%
  %5 citations counted in INSPIRE as of 02 Dec 2015

%\cite{Arbelaez:2015ila}
\bibitem{Arbelaez:2015ila}
  C.~Arbelaez, R.~Longas, D.~Restrepo and O.~Zapata,
  %``Fermion dark matter from SO(10),''
  arXiv:1509.06313 [hep-ph].
  %%CITATION = ARXIV:1509.06313;%%

%\cite{Boucenna:2015sdg}
\bibitem{Boucenna:2015sdg}
  S.~M.~Boucenna, M.~B.~Krauss and E.~Nardi,
  %``Dark Matter from the vector of SO(10),''
  arXiv:1511.02524 [hep-ph].
  %%CITATION = ARXIV:1511.02524;%%

%\cite{Farrar:1978xj}
\bibitem{Farrar:1978xj}
  G.~R.~Farrar and P.~Fayet,
  %``Phenomenology of the Production, Decay, and Detection of New Hadronic States Associated with Supersymmetry,''
  Phys.\ Lett.\ B {\bf 76}, 575 (1978);
  %%CITATION = PHLTA,B76,575;%%
  %1033 citations counted in INSPIRE as of 26 gen 2015
%
%\cite{Dimopoulos:1981zb}
%\bibitem{Dimopoulos:1981zb}
  S.~Dimopoulos and H.~Georgi,
  %``Softly Broken Supersymmetry and SU(5),''
  Nucl.\ Phys.\ B {\bf 193}, 150 (1981);
  %%CITATION = NUPHA,B193,150;%%
  %2142 citations counted in INSPIRE as of 26 Jan 2015
%
%\cite{Weinberg:1981wj}
%\bibitem{Weinberg:1981wj}
  S.~Weinberg,
  %``Supersymmetry at Ordinary Energies. 1. Masses and Conservation Laws,''
  Phys.\ Rev.\ D {\bf 26}, 287 (1982);
  %%CITATION = PHRVA,D26,287;%%
  %925 citations counted in INSPIRE as of 26 Jan 2015
%
%\cite{Sakai:1981pk}
%\bibitem{Sakai:1981pk}
  N.~Sakai and T.~Yanagida,
  %``Proton Decay in a Class of Supersymmetric Grand Unified Models,''
  Nucl.\ Phys.\ B {\bf 197}, 533 (1982);
  %%CITATION = NUPHA,B197,533;%%
  %750 citations counted in INSPIRE as of 26 gen 2015
%
%\cite{Dimopoulos:1981dw}
%\bibitem{Dimopoulos:1981dw}
  S.~Dimopoulos, S.~Raby and F.~Wilczek,
  %``Proton Decay in Supersymmetric Models,''
  Phys.\ Lett.\ B {\bf 112}, 133 (1982).
  %%CITATION = PHLTA,B112,133;%%
  %356 citations counted in INSPIRE as of 26 gen 2015

%\cite{Aydemir:2015nfa}
\bibitem{Aydemir:2015nfa}
  U.~Aydemir, D.~Minic, C.~Sun and T.~Takeuchi,
  %``Pati-Salam Unification from Non-commutative Geometry and the TeV-scale $W_R$ boson,''
  arXiv:1509.01606 [hep-ph].
  %%CITATION = ARXIV:1509.01606;%%
  %4 citations counted in INSPIRE as of 02 Dec 2015

%\cite{Bandyopadhyay:2015fka}
\bibitem{Bandyopadhyay:2015fka}
  T.~Bandyopadhyay, B.~Brahmachari and A.~Raychaudhuri,
  %``Implications of the CMS search for W_R on Grand Unification,''
  arXiv:1509.03232 [hep-ph].
  %%CITATION = ARXIV:1509.03232;%%
  %4 citations counted in INSPIRE as of 02 Dec 2015

%\cite{Aydemir:2015oob}
\bibitem{Aydemir:2015oob}
  U.~Aydemir,
  %``SO(10) Grand Unification in light of recent LHC searches and colored scalars at the TeV-scale,''
  arXiv:1512.00568 [hep-ph].
  %%CITATION = ARXIV:1512.00568;%%

%\cite{Davidson:1993qk}
\bibitem{Davidson:1993qk}
  S.~Davidson, D.~C.~Bailey and B.~A.~Campbell,
  %``Model independent constraints on leptoquarks from rare processes,''
  Z.\ Phys.\ C {\bf 61}, 613 (1994)
  doi:10.1007/BF01552629
  [hep-ph/9309310].
  %%CITATION = doi:10.1007/BF01552629;%%
  %349 citations counted in INSPIRE as of 02 Dec 2015

%\cite{Kuzmin:1980yp}
\bibitem{Kuzmin:1980yp}
  V.~A.~Kuzmin and M.~E.~Shaposhnikov,
  %``Baryon Asymmetry of the Universe Versus Left-right Symmetry,''
  Phys.\ Lett.\ B {\bf 92}, 115 (1980);
  %%CITATION = PHLTA,B92,115;%%
  %67 citations counted in INSPIRE as of 26 Jan 2015
%
%\cite{Kibble:1982dd}
%\bibitem{Kibble:1982dd}
  T.~W.~B.~Kibble, G.~Lazarides and Q.~Shafi,
  %``Walls Bounded by Strings,''
  Phys.\ Rev.\ D {\bf 26}, 435 (1982);
  %%CITATION = PHRVA,D26,435;%%
  %164 citations counted in INSPIRE as of 26 Jan 2015
%
%\cite{Chang:1983fu}
%\bibitem{Chang:1983fu}
  D.~Chang, R.~N.~Mohapatra and M.~K.~Parida,
  %``Decoupling Parity and SU(2)-R Breaking Scales: A New Approach to Left-Right Symmetric Models,''
  Phys.\ Rev.\ Lett.\  {\bf 52}, 1072 (1984);
  %%CITATION = PRLTA,52,1072;%%
  %204 citations counted in INSPIRE as of 26 gen 2015
%
%\cite{Chang:1984uy}
%\bibitem{Chang:1984uy}
  D.~Chang, R.~N.~Mohapatra and M.~K.~Parida,
  %``A New Approach to Left-Right Symmetry Breaking in Unified Gauge Theories,''
  Phys.\ Rev.\ D {\bf 30}, 1052 (1984);
  %%CITATION = PHRVA,D30,1052;%%
  %164 citations counted in INSPIRE as of 26 Jan 2015
%
%\cite{Chang:1984qr}
%\bibitem{Chang:1984qr}
  D.~Chang, R.~N.~Mohapatra, J.~Gipson, R.~E.~Marshak and M.~K.~Parida,
  %``Experimental Tests of New SO(10) Grand Unification,''
  Phys.\ Rev.\ D {\bf 31}, 1718 (1985).
  %%CITATION = PHRVA,D31,1718;%%
  %157 citations counted in INSPIRE as of 26 Jan 2015

%\cite{Pati:1974yy}
\bibitem{Pati:1974yy}
  J.~C.~Pati and A.~Salam,
  %``Lepton Number as the Fourth Color,''
  Phys.\ Rev.\ D {\bf 10}, 275 (1974)
  [Phys.\ Rev.\ D {\bf 11}, 703 (1975)];
  %%CITATION = PHRVA,D10,275;%%
  %3832 citations counted in INSPIRE as of 21 sept. 2015
%\cite{Mohapatra:1974hk}
%\bibitem{Mohapatra:1974hk}
  R.~N.~Mohapatra and J.~C.~Pati,
  %``Left-Right Gauge Symmetry and an Isoconjugate Model of CP Violation,''
  Phys.\ Rev.\ D {\bf 11}, 566 (1975);
  %%CITATION = PHRVA,D11,566;%%
  %1727 citations counted in INSPIRE as of 21 sept. 2015
%\cite{Mohapatra:1974gc}
%\bibitem{Mohapatra:1974gc}
  R.~N.~Mohapatra and J.~C.~Pati,
  %``A Natural Left-Right Symmetry,''
  Phys.\ Rev.\ D {\bf 11}, 2558 (1975);
  %%CITATION = PHRVA,D11,2558;%%
  %1198 citations counted in INSPIRE as of 21 sept. 2015
%\cite{Senjanovic:1975rk}
%\bibitem{Senjanovic:1975rk}
  G.~Senjanovic and R.~N.~Mohapatra,
  %``Exact Left-Right Symmetry and Spontaneous Violation of Parity,''
  Phys.\ Rev.\ D {\bf 12}, 1502 (1975).
  %%CITATION = PHRVA,D12,1502;%%
  %1614 citations counted in INSPIRE as of 21 sept. 2015

%\cite{Ade:2015xua}
\bibitem{Ade:2015xua}
  P.~A.~R.~Ade {\it et al.} [Planck Collaboration],
  %``Planck 2015 results. XIII. Cosmological parameters,''
  arXiv:1502.01589 [astro-ph.CO].
  %%CITATION = ARXIV:1502.01589;%%
  %948 citations counted in INSPIRE as of 02 Dec 2015

%\cite{Cirelli:2005uq}
\bibitem{Cirelli:2005uq}
  M.~Cirelli, N.~Fornengo and A.~Strumia,
  %``Minimal dark matter,''
  Nucl.\ Phys.\ B {\bf 753}, 178 (2006)
  [hep-ph/0512090];
  %%CITATION = HEP-PH/0512090;%%
  %302 citations counted in INSPIRE as of 23 Nov 2014
%
%\cite{Cirelli:2007xd}
%\bibitem{Cirelli:2007xd}
  M.~Cirelli, A.~Strumia and M.~Tamburini,
  %``Cosmology and Astrophysics of Minimal Dark Matter,''
  Nucl.\ Phys.\ B {\bf 787}, 152 (2007)
  [arXiv:0706.4071 [hep-ph]];
  %%CITATION = ARXIV:0706.4071;%%
  %208 citations counted in INSPIRE as of 23 Nov 2014
%
%
%\cite{Cirelli:2009uv}
%\bibitem{Cirelli:2009uv}
  M.~Cirelli and A.~Strumia,
  %``Minimal Dark Matter: Model and results,''
  New J.\ Phys.\  {\bf 11}, 105005 (2009)
  [arXiv:0903.3381 [hep-ph]].
  %%CITATION = ARXIV:0903.3381;%%
  %61 citations counted in INSPIRE as of 23 Nov 2014

%\cite{Hisano:2003ec}
\bibitem{Hisano:2003ec}
  J.~Hisano, S.~Matsumoto and M.~M.~Nojiri,
  %``Explosive dark matter annihilation,''
  Phys.\ Rev.\ Lett.\  {\bf 92}, 031303 (2004)
  [hep-ph/0307216];
  %%CITATION = HEP-PH/0307216;%%
  %250 citations counted in INSPIRE as of 14 Jun 2015
%\cite{Hisano:2004ds}
%\bibitem{Hisano:2004ds}
  J.~Hisano, S.~Matsumoto, M.~M.~Nojiri and O.~Saito,
  %``Non-perturbative effect on dark matter annihilation and gamma ray signature from galactic center,''
  Phys.\ Rev.\ D {\bf 71}, 063528 (2005)
  [hep-ph/0412403].
  %%CITATION = HEP-PH/0412403;%%
  %297 citations counted in INSPIRE as of 14 juin 2015

%\cite{Hisano:2006nn}
\bibitem{Hisano:2006nn}
  J.~Hisano, S.~Matsumoto, M.~Nagai, O.~Saito and M.~Senami,
  %``Non-perturbative effect on thermal relic abundance of dark matter,''
  Phys.\ Lett.\ B {\bf 646}, 34 (2007)
%  doi:10.1016/j.physletb.2007.01.012
  [hep-ph/0610249].
  %%CITATION = doi:10.1016/j.physletb.2007.01.012;%%
  %212 citations counted in INSPIRE as of 03 Dec 2015

%\cite{Aad:2014cka}
\bibitem{Aad:2014cka}
  G.~Aad {\it et al.} [ATLAS Collaboration],
  %``Search for high-mass dilepton resonances in pp collisions at $\sqrt{s}=8$  TeV with the ATLAS detector,''
  Phys.\ Rev.\ D {\bf 90}, no. 5, 052005 (2014)
  [arXiv:1405.4123 [hep-ex]].
  %%CITATION = ARXIV:1405.4123;%%
  %126 citations counted in INSPIRE as of 29 Sep 2015

%\cite{Khachatryan:2014fba}
\bibitem{Khachatryan:2014fba}
  V.~Khachatryan {\it et al.} [CMS Collaboration],
  %``Search for physics beyond the standard model in dilepton mass spectra in proton-proton collisions at $ \sqrt{s}=8 $ TeV,''
  JHEP {\bf 1504}, 025 (2015)
  [arXiv:1412.6302 [hep-ex]].
  %%CITATION = ARXIV:1412.6302;%%
  %52 citations counted in INSPIRE as of 29 Sep 2015

%\cite{Accomando:2010fz}
\bibitem{Accomando:2010fz}
  E.~Accomando, A.~Belyaev, L.~Fedeli, S.~F.~King and C.~Shepherd-Themistocleous,
  %``Z' physics with early LHC data,''
  Phys.\ Rev.\ D {\bf 83}, 075012 (2011)
%  doi:10.1103/PhysRevD.83.075012
  [arXiv:1010.6058 [hep-ph]].
  %%CITATION = doi:10.1103/PhysRevD.83.075012;%%
  %83 citations counted in INSPIRE as of 04 Dec 2015

%\cite{Godfrey:2013eta}
\bibitem{Godfrey:2013eta}
  S.~Godfrey and T.~Martin,
  %``Z' Discovery Reach at Future Hadron Colliders: A Snowmass White Paper,''
  arXiv:1309.1688 [hep-ph].
  %%CITATION = ARXIV:1309.1688;%%
  %14 citations counted in INSPIRE as of 04 Dec 2015

\bibitem{CMS-DP-2015-039}
 CMS Collaboration, CMS-DP-2015-039 (2015).

%\cite{Bjorken:1977vt}
\bibitem{Bjorken:1977vt}
  J.~D.~Bjorken and S.~Weinberg,
  %``A Mechanism for Nonconservation of Muon Number,''
  Phys.\ Rev.\ Lett.\  {\bf 38}, 622 (1977).
%  doi:10.1103/PhysRevLett.38.622
  %%CITATION = doi:10.1103/PhysRevLett.38.622;%%
  %147 citations counted in INSPIRE as of 03 Dec 2015

%\cite{McWilliams:1980kj}
\bibitem{McWilliams:1980kj}
  B.~McWilliams and L.~F.~Li,
  %``Virtual Effects of Higgs Particles,''
  Nucl.\ Phys.\ B {\bf 179}, 62 (1981).
%  doi:10.1016/0550-3213(81)90249-2
  %%CITATION = doi:10.1016/0550-3213(81)90249-2;%%
  %111 citations counted in INSPIRE as of 03 Dec 2015

%\cite{Shanker:1981mj}
\bibitem{Shanker:1981mj}
  O.~U.~Shanker,
  %``Flavor Violation, Scalar Particles and Leptoquarks,''
  Nucl.\ Phys.\ B {\bf 206}, 253 (1982).
%  doi:10.1016/0550-3213(82)90534-X
  %%CITATION = doi:10.1016/0550-3213(82)90534-X;%%
  %133 citations counted in INSPIRE as of 03 Dec 2015


%\cite{Khachatryan:2015kon}
\bibitem{Khachatryan:2015kon}
  V.~Khachatryan {\it et al.} [CMS Collaboration],
  %``Search for lepton-flavour-violating decays of the Higgs boson,''
  Phys.\ Lett.\ B {\bf 749}, 337 (2015)
  [arXiv:1502.07400 [hep-ex]].
  %%CITATION = ARXIV:1502.07400;%%
  %45 citations counted in INSPIRE as of 27 Sep 2015

%\cite{Sierra:2014nqa}
\bibitem{Sierra:2014nqa}
  D.~Aristizabal Sierra and A.~Vicente,
  %``Explaining the CMS Higgs flavor violating decay excess,''
  Phys.\ Rev.\ D {\bf 90}, no. 11, 115004 (2014)
  [arXiv:1409.7690 [hep-ph]];
  %%CITATION = ARXIV:1409.7690;%%
  %28 citations counted in INSPIRE as of 27 Sep 2015
%\cite{Crivellin:2015mga}
%\bibitem{Crivellin:2015mga}
  A.~Crivellin, G.~D'Ambrosio and J.~Heeck,
  %``Explaining $h\to\mu^\pm\tau^\mp$, $B\to K^* \mu^+\mu^-$ and $B\to K \mu^+\mu^-/B\to K e^+e^-$ in a two-Higgs-doublet model with gauged $L_\mu-L_\tau$,''
  Phys.\ Rev.\ Lett.\  {\bf 114}, 151801 (2015)
  [arXiv:1501.00993 [hep-ph]];
  %%CITATION = ARXIV:1501.00993;%%
  %53 citations counted in INSPIRE as of 27 Sep 2015
%\cite{deLima:2015pqa}
%\bibitem{deLima:2015pqa}
  L.~de Lima, C.~S.~Machado, R.~D.~Matheus and L.~A.~F.~do Prado,
  %``Higgs Flavor Violation as a Signal to Discriminate Models,''
  arXiv:1501.06923 [hep-ph];
  %%CITATION = ARXIV:1501.06923;%%
  %13 citations counted in INSPIRE as of 27 Sep 2015
%\cite{Dorsner:2015mja}
%\bibitem{Dorsner:2015mja}
  I.~Dor\v{s}ner, S.~Fajfer, A.~Greljo, J.~F.~Kamenik, N.~Ko\v{s}nik and I.~Ni\v{s}and\v{z}i\'{c},
  %``New Physics Models Facing Lepton Flavor Violating Higgs Decays at the Percent Level,''
  JHEP {\bf 1506}, 108 (2015)
  [arXiv:1502.07784 [hep-ph]];
  %%CITATION = ARXIV:1502.07784;%%
  %19 citations counted in INSPIRE as of 27 Sep 2015
%\cite{Omura:2015nja}
%\bibitem{Omura:2015nja}
  Y.~Omura, E.~Senaha and K.~Tobe,
  %``Lepton-flavor-violating Higgs decay $h \to \mu\tau$ and muon anomalous magnetic moment in a general two Higgs doublet model,''
  JHEP {\bf 1505}, 028 (2015)
  [arXiv:1502.07824 [hep-ph]];
  %%CITATION = ARXIV:1502.07824;%%
  %12 citations counted in INSPIRE as of 27 Sep 2015
%\cite{Das:2015zwa}
%\bibitem{Das:2015zwa}
  D.~Das and A.~Kundu,
  %``Two hidden scalars around 125 GeV and h→μτ,''
  Phys.\ Rev.\ D {\bf 92}, no. 1, 015009 (2015)
 % doi:10.1103/PhysRevD.92.015009
  [arXiv:1504.01125 [hep-ph]];
  %%CITATION = doi:10.1103/PhysRevD.92.015009;%%
  %8 citations counted in INSPIRE as of 02 Dec 2015
%\cite{Mao:2015hwa}
%\bibitem{Mao:2015hwa}
  Y.~n.~Mao and S.~h.~Zhu,
  %``On the Higgs-$\mu$-$\tau$ Coupling at High and Low Energy Colliders,''
  arXiv:1505.07668 [hep-ph];
  %%CITATION = ARXIV:1505.07668;%%
  %8 citations counted in INSPIRE as of 02 Dec 2015
%\cite{Crivellin:2015hha}
%\bibitem{Crivellin:2015hha}
  A.~Crivellin, J.~Heeck and P.~Stoffer,
  %``A perturbed lepton-specific two-Higgs-doublet model facing experimental hints for physics beyond the Standard Model,''
  arXiv:1507.07567 [hep-ph];
  %%CITATION = ARXIV:1507.07567;%%
  %13 citations counted in INSPIRE as of 02 Dec 2015
%\cite{Goto:2015iha}
%\bibitem{Goto:2015iha}
  T.~Goto, R.~Kitano and S.~Mori,
  %``Lepton flavor violating $Z$-boson couplings from nonstandard Higgs interactions,''
  Phys.\ Rev.\ D {\bf 92}, no. 7, 075021 (2015)
%  doi:10.1103/PhysRevD.92.075021
  [arXiv:1507.03234 [hep-ph]];
  %%CITATION = doi:10.1103/PhysRevD.92.075021;%%
  %4 citations counted in INSPIRE as of 02 Dec 2015
%\cite{Chiang:2015cba}
%\bibitem{Chiang:2015cba}
  C.~W.~Chiang, H.~Fukuda, M.~Takeuchi and T.~T.~Yanagida,
  %``Flavor-Changing Neutral-Current Decays in Top-Specific Variant Axion Model,''
  JHEP {\bf 1511}, 057 (2015)
 % doi:10.1007/JHEP11(2015)057
  [arXiv:1507.04354 [hep-ph]];
  %%CITATION = doi:10.1007/JHEP11(2015)057;%%
  %7 citations counted in INSPIRE as of 02 Dec 2015
%\cite{Crivellin:2015hha}
%\bibitem{Crivellin:2015hha}
  A.~Crivellin, J.~Heeck and P.~Stoffer,
  %``A perturbed lepton-specific two-Higgs-doublet model facing experimental hints for physics beyond the Standard Model,''
  arXiv:1507.07567 [hep-ph];
  %%CITATION = ARXIV:1507.07567;%%
  %13 citations counted in INSPIRE as of 02 Dec 2015
%\cite{Botella:2015hoa}
%\bibitem{Botella:2015hoa}
  F.~J.~Botella, G.~C.~Branco, M.~Nebot and M.~N.~Rebelo,
  %``Flavour Changing Higgs Couplings in a Class of Two Higgs Doublet Models,''
  arXiv:1508.05101 [hep-ph];
  %%CITATION = ARXIV:1508.05101;%%
  %10 citations counted in INSPIRE as of 02 Dec 2015
%\cite{Omura:2015xcg}
%\bibitem{Omura:2015xcg}
  Y.~Omura, E.~Senaha and K.~Tobe,
  %``$\tau$- and $\mu$-physics in a general two Higgs doublet model with $\mu-\tau$ flavor violation,''
  arXiv:1511.08880 [hep-ph].
  %%CITATION = ARXIV:1511.08880;%%

%\cite{Minkowski:1977sc}
\bibitem{Minkowski:1977sc}
  P.~Minkowski,
  %``mu --> e gamma at a Rate of One Out of 1-Billion Muon Decays?,''
  Phys.\ Lett.\ B {\bf 67}, 421 (1977);
  %%CITATION = PHLTA,B67,421;%%
  %2028 citations counted in INSPIRE as of 08 Feb 2015
%\cite{Yanagida:1979as}
%\bibitem{Yanagida:1979as}
  T.~Yanagida,
  %``Horizontal Symmetry And Masses Of Neutrinos,''
  Conf.\ Proc.\ C {\bf 7902131}, 95 (1979);
  %%CITATION = CONFP,C7902131,95;%%
  %977 citations counted in INSPIRE as of 08 Feb 2015
%\cite{GellMann:1980vs}
%\bibitem{GellMann:1980vs}
  M.~Gell-Mann, P.~Ramond and R.~Slansky,
  %``Complex Spinors and Unified Theories,''
  Conf.\ Proc.\ C {\bf 790927}, 315 (1979)
  [arXiv:1306.4669 [hep-th]];
  %%CITATION = ARXIV:1306.4669;%%
  %2032 citations counted in INSPIRE as of 08 Feb 2015
%\cite{Glashow:1979nm}
%\bibitem{Glashow:1979nm}
  S.~L.~Glashow,
  %``The Future of Elementary Particle Physics,''
  NATO Sci.\ Ser.\ B {\bf 59}, 687 (1980);
  %304 citations counted in INSPIRE as of 08 Feb 2015
%\cite{Mohapatra:1979ia}
%\bibitem{Mohapatra:1979ia}
  R.~N.~Mohapatra and G.~Senjanovic,
  %``Neutrino Mass and Spontaneous Parity Violation,''
  Phys.\ Rev.\ Lett.\  {\bf 44}, 912 (1980);
  %%CITATION = PRLTA,44,912;%%
  %3578 citations counted in INSPIRE as of 08 Feb 2015
  R.~N.~Mohapatra and G.~Senjanovic,
  %``Neutrino Masses and Mixings in Gauge Models with Spontaneous Parity Violation,''
  Phys.\ Rev.\ D {\bf 23}, 165 (1981).
  %%CITATION = PHRVA,D23,165;%%

%\cite{Babu:1992ia}
\bibitem{Babu:1992ia}
  K.~S.~Babu and R.~N.~Mohapatra,
  %``Predictive neutrino spectrum in minimal SO(10) grand unification,''
  Phys.\ Rev.\ Lett.\  {\bf 70}, 2845 (1993)
  [hep-ph/9209215];
  %%CITATION = HEP-PH/9209215;%%
  %338 citations counted in INSPIRE as of 28 Sep 2015
%\cite{Matsuda:2000zp}
%\bibitem{Matsuda:2000zp}
  K.~Matsuda, Y.~Koide and T.~Fukuyama,
  %``Can the SO(10) model with two Higgs doublets reproduce the observed fermion masses?,''
  Phys.\ Rev.\ D {\bf 64}, 053015 (2001)
  [hep-ph/0010026];
  %%CITATION = HEP-PH/0010026;%%
  %91 citations counted in INSPIRE as of 28 Sep 2015
%\cite{Bajc:2005zf}
%\bibitem{Bajc:2005zf}
  B.~Bajc, A.~Melfo, G.~Senjanovic and F.~Vissani,
  %``Yukawa sector in non-supersymmetric renormalizable SO(10),''
  Phys.\ Rev.\ D {\bf 73}, 055001 (2006)
  [hep-ph/0510139];
  %%CITATION = HEP-PH/0510139;%%
  %55 citations counted in INSPIRE as of 28 Sep 2015
%\cite{Fukuyama:2015kra}
%\bibitem{Fukuyama:2015kra}
  T.~Fukuyama, K.~Ichikawa and Y.~Mimura,
  %``Revisiting fermion mass and mixing fits in the minimal SUSY $SO(10)$ GUT,''
  arXiv:1508.07078 [hep-ph].
  %%CITATION = ARXIV:1508.07078;%%
  %1 citations counted in INSPIRE as of 28 Sep 2015

%\cite{Zhang:2007da}
\bibitem{Zhang:2007da}
  Y.~Zhang, H.~An, X.~Ji and R.~N.~Mohapatra,
  %``General CP Violation in Minimal Left-Right Symmetric Model and Constraints on the Right-Handed Scale,''
  Nucl.\ Phys.\ B {\bf 802}, 247 (2008)
%  doi:10.1016/j.nuclphysb.2008.05.019
  [arXiv:0712.4218 [hep-ph]];
  %%CITATION = doi:10.1016/j.nuclphysb.2008.05.019;%%
  %103 citations counted in INSPIRE as of 03 Dec 2015
%\cite{Maiezza:2010ic}
%\bibitem{Maiezza:2010ic}
  A.~Maiezza, M.~Nemevsek, F.~Nesti and G.~Senjanovic,
  %``Left-Right Symmetry at LHC,''
  Phys.\ Rev.\ D {\bf 82}, 055022 (2010)
 % doi:10.1103/PhysRevD.82.055022
  [arXiv:1005.5160 [hep-ph]];
  %%CITATION = doi:10.1103/PhysRevD.82.055022;%%
  %137 citations counted in INSPIRE as of 03 Dec 2015
%\cite{Bertolini:2014sua}
%\bibitem{Bertolini:2014sua}
  S.~Bertolini, A.~Maiezza and F.~Nesti,
  %``Present and Future K and B Meson Mixing Constraints on TeV Scale Left-Right Symmetry,''
  Phys.\ Rev.\ D {\bf 89}, no. 9, 095028 (2014)
 % doi:10.1103/PhysRevD.89.095028
  [arXiv:1403.7112 [hep-ph]].
  %%CITATION = doi:10.1103/PhysRevD.89.095028;%%
  %34 citations counted in INSPIRE as of 03 Dec 2015

%\cite{Goncalves:2015yua}
\bibitem{Goncalves:2015yua}
  D.~Gon\c{c}alves, F.~Krauss and M.~Spannowsky,
  %``Augmenting the diboson excess for the LHC Run II,''
  Phys.\ Rev.\ D {\bf 92}, no. 5, 053010 (2015)
%  doi:10.1103/PhysRevD.92.053010
  [arXiv:1508.04162 [hep-ph]].
  %%CITATION = doi:10.1103/PhysRevD.92.053010;%%
  %14 citations counted in INSPIRE as of 03 Dec 2015

%\cite{Aad:2015rpa}
\bibitem{Aad:2015rpa}
  G.~Aad {\it et al.} [ATLAS Collaboration],
  %``Identification of Boosted, Hadronically Decaying W Bosons and Comparisons with ATLAS Data Taken at $\sqrt{s} = 8$ TeV,''
  arXiv:1510.05821 [hep-ex].
  %%CITATION = ARXIV:1510.05821;%%
  %2 citations counted in INSPIRE as of 03 Dec 2015


%\cite{Khachatryan:2014hpa}
\bibitem{Khachatryan:2014hpa}
  V.~Khachatryan {\it et al.} [CMS Collaboration],
  %``Search for massive resonances in dijet systems containing jets tagged as W or Z boson decays in pp collisions at $ \sqrt{s} $ = 8 TeV,''
  JHEP {\bf 1408}, 173 (2014)
  [arXiv:1405.1994 [hep-ex]].
  %%CITATION = ARXIV:1405.1994;%%
  %80 citations counted in INSPIRE as of 16 sept. 2015

\bibitem{ATLAS-CONF-2015-045}
ATLAS Collaboration, ATLAS-CONF-2015-045.
% http://cds.cern.ch/record/2052583

%\cite{Aad:2014pha}
\bibitem{Aad:2014pha}
  G.~Aad {\it et al.} [ATLAS Collaboration],
  %``Search for $WZ$ resonances in the fully leptonic channel using $pp$ collisions at $\sqrt{s}$ = 8 TeV with the ATLAS detector,''
  Phys.\ Lett.\ B {\bf 737}, 223 (2014)
  [arXiv:1406.4456 [hep-ex]].
  %%CITATION = ARXIV:1406.4456;%%
  %34 citations counted in INSPIRE as of 16 sept. 2015

%\cite{Aad:2014xka}
\bibitem{Aad:2014xka}
  G.~Aad {\it et al.} [ATLAS Collaboration],
  %``Search for resonant diboson production in the $\mathrm {\ell \ell }q\bar{q}$ final state in $pp$ collisions at $\sqrt{s} = 8$ TeV with the ATLAS detector,''
  Eur.\ Phys.\ J.\ C {\bf 75}, no. 2, 69 (2015)
  [arXiv:1409.6190 [hep-ex]].
  %%CITATION = ARXIV:1409.6190;%%
  %28 citations counted in INSPIRE as of 16 sept. 2015

%\cite{Aad:2015ufa}
\bibitem{Aad:2015ufa}
  G.~Aad {\it et al.} [ATLAS Collaboration],
  %``Search for production of $WW/WZ$ resonances decaying to a lepton, neutrino and jets in $pp$ collisions at $\sqrt{s}=8$  TeV with the ATLAS detector,''
  Eur.\ Phys.\ J.\ C {\bf 75}, no. 5, 209 (2015)
  [Eur.\ Phys.\ J.\ C {\bf 75}, 370 (2015)]
  [arXiv:1503.04677 [hep-ex]].
  %%CITATION = ARXIV:1503.04677;%%
  %27 citations counted in INSPIRE as of 16 sept. 2015

%\cite{Khachatryan:2014gha}
\bibitem{Khachatryan:2014gha}
  V.~Khachatryan {\it et al.} [CMS Collaboration],
  %``Search for massive resonances decaying into pairs of boosted bosons in semi-leptonic final states at $\sqrt{s} =$ 8 TeV,''
  JHEP {\bf 1408}, 174 (2014)
  [arXiv:1405.3447 [hep-ex]].
  %%CITATION = ARXIV:1405.3447;%%
  %70 citations counted in INSPIRE as of 16 sept. 2015

%\cite{Khachatryan:2014xja}
\bibitem{Khachatryan:2014xja}
  V.~Khachatryan {\it et al.} [CMS Collaboration],
  %``Search for new resonances decaying via WZ to leptons in proton-proton collisions at $\sqrt s =$ 8 TeV,''
  Phys.\ Lett.\ B {\bf 740}, 83 (2015)
  [arXiv:1407.3476 [hep-ex]].
  %%CITATION = ARXIV:1407.3476;%%
  %29 citations counted in INSPIRE as of 16 sept. 2015

%\cite{Khachatryan:2015sja}
\bibitem{Khachatryan:2015sja}
  V.~Khachatryan {\it et al.} [CMS Collaboration],
  %``Search for resonances and quantum black holes using dijet mass spectra in proton-proton collisions at $\sqrt{s} =$ 8 TeV,''
  Phys.\ Rev.\ D {\bf 91}, no. 5, 052009 (2015)
  [arXiv:1501.04198 [hep-ex]].
  %%CITATION = ARXIV:1501.04198;%%
  %53 citations counted in INSPIRE as of 16 sept. 2015

%\cite{Aad:2014aqa}
\bibitem{Aad:2014aqa}
  G.~Aad {\it et al.} [ATLAS Collaboration],
  %``Search for new phenomena in the dijet mass distribution using $p-p$ collision data at $\sqrt{s}=8$ TeV with the ATLAS detector,''
  Phys.\ Rev.\ D {\bf 91}, no. 5, 052007 (2015)
  [arXiv:1407.1376 [hep-ex]].
  %%CITATION = ARXIV:1407.1376;%%
  %69 citations counted in INSPIRE as of 16 sept. 2015

%\cite{CMS:2015gla}
\bibitem{CMS:2015gla}
  CMS Collaboration [CMS Collaboration],
  %``Search for massive WH resonances decaying to $\ell \nu {\rm b \bar{b}}$ final state in the boosted regime at $\sqrt{s}=8$\,TeV,''
  CMS-PAS-EXO-14-010.
  %%CITATION = CMS-PAS-EXO-14-010;%%
  %18 citations counted in INSPIRE as of 16 sept. 2015

%\cite{Aad:2015yza}
\bibitem{Aad:2015yza}
  G.~Aad {\it et al.} [ATLAS Collaboration],
  %``Search for a new resonance decaying to a W or Z boson and a Higgs boson in the $\ell \ell / \ell \nu / \nu \nu + b \bar{b}$ final states with the ATLAS detector,''
  Eur.\ Phys.\ J.\ C {\bf 75}, no. 6, 263 (2015)
  [arXiv:1503.08089 [hep-ex]].
  %%CITATION = ARXIV:1503.08089;%%
  %16 citations counted in INSPIRE as of 16 sept. 2015

%\cite{Khachatryan:2015bma}
\bibitem{Khachatryan:2015bma}
  V.~Khachatryan {\it et al.} [CMS Collaboration],
  %``Search for A Massive Resonance Decaying into a Higgs Boson and a W or Z Boson in Hadronic Final States in Proton-Proton Collisions at $\sqrt{s}$ = 8 TeV,''
  arXiv:1506.01443 [hep-ex].
  %%CITATION = ARXIV:1506.01443;%%
  %22 citations counted in INSPIRE as of 16 sept. 2015

%\cite{Aad:2014xra}
\bibitem{Aad:2014xra}
  G.~Aad {\it et al.} [ATLAS Collaboration],
  %``Search for $W' \rightarrow tb \rightarrow qqbb$ decays in $pp$ collisions at $\sqrt{s}$  = 8 TeV with the ATLAS detector,''
  Eur.\ Phys.\ J.\ C {\bf 75}, no. 4, 165 (2015)
  [arXiv:1408.0886 [hep-ex]].
  %%CITATION = ARXIV:1408.0886;%%
  %21 citations counted in INSPIRE as of 16 sept. 2015

%\cite{Aad:2014xea}
\bibitem{Aad:2014xea}
  G.~Aad {\it et al.} [ATLAS Collaboration],
  %``Search for $W' \to t\bar{b}$ in the lepton plus jets final state in proton-proton collisions at a centre-of-mass energy of $\sqrt{s}$ = 8 TeV with the ATLAS detector,''
  Phys.\ Lett.\ B {\bf 743}, 235 (2015)
  [arXiv:1410.4103 [hep-ex]].
  %%CITATION = ARXIV:1410.4103;%%
  %20 citations counted in INSPIRE as of 16 sept. 2015

%\cite{Chatrchyan:2014koa}
\bibitem{Chatrchyan:2014koa}
  S.~Chatrchyan {\it et al.} [CMS Collaboration],
  %``Search for W' $\to $ tb decays in the lepton + jets final state in pp collisions at $\sqrt{s}$ = 8 TeV,''
  JHEP {\bf 1405}, 108 (2014)
  [arXiv:1402.2176 [hep-ex]].
  %%CITATION = ARXIV:1402.2176;%%
  %36 citations counted in INSPIRE as of 16 sept. 2015

%\cite{Khachatryan:2015edz}
\bibitem{Khachatryan:2015edz}
  V.~Khachatryan {\it et al.} [CMS Collaboration],
  %``Search for W' to tb in proton-proton collisions at sqrt(s) = 8 TeV,''
  arXiv:1509.06051 [hep-ex].
  %%CITATION = ARXIV:1509.06051;%%


%\cite{Alwall:2014hca}
\bibitem{Alwall:2014hca}
  J.~Alwall {\it et al.},
  %``The automated computation of tree-level and next-to-leading order differential cross sections, and their matching to parton shower simulations,''
  JHEP {\bf 1407}, 079 (2014)
  [arXiv:1405.0301 [hep-ph]].
  %%CITATION = ARXIV:1405.0301;%%
  %485 citations counted in INSPIRE as of 20 sept. 2015

%\cite{Cao:2012ng}
\bibitem{Cao:2012ng}
  Q.~H.~Cao, Z.~Li, J.~H.~Yu and C.~P.~Yuan,
  %``Discovery and Identification of W' and Z' in SU(2) x SU(2) x U(1) Models at the LHC,''
  Phys.\ Rev.\ D {\bf 86}, 095010 (2012)
  [arXiv:1205.3769 [hep-ph]].
  %%CITATION = ARXIV:1205.3769;%%
  %31 citations counted in INSPIRE as of 20 sept. 2015

%\cite{Jezo:2014wra}
\bibitem{Jezo:2014wra}
  T.~Jezo, M.~Klasen, D.~R.~Lamprea, F.~Lyonnet and I.~Schienbein,
  %``NLO+NLL limits on $W'$ and $Z'$ gauge boson masses in general extensions of the Standard Model,''
  JHEP {\bf 1412}, 092 (2014)
  [arXiv:1410.4692 [hep-ph]].
  %%CITATION = ARXIV:1410.4692;%%
  %3 citations counted in INSPIRE as of 20 sept. 2015

%\cite{Grojean:2011vu}
\bibitem{Grojean:2011vu}
  C.~Grojean, E.~Salvioni and R.~Torre,
  %``A weakly constrained W' at the early LHC,''
  JHEP {\bf 1107}, 002 (2011)
  doi:10.1007/JHEP07(2011)002
  [arXiv:1103.2761 [hep-ph]].
  %%CITATION = doi:10.1007/JHEP07(2011)002;%%
  %43 citations counted in INSPIRE as of 03 Dec 2015

%\cite{Peskin:1990zt}
\bibitem{Peskin:1990zt}
  M.~E.~Peskin and T.~Takeuchi,
  %``A New constraint on a strongly interacting Higgs sector,''
  Phys.\ Rev.\ Lett.\  {\bf 65}, 964 (1990);
%  doi:10.1103/PhysRevLett.65.964
  %%CITATION = doi:10.1103/PhysRevLett.65.964;%%
  %1536 citations counted in INSPIRE as of 03 Dec 2015
%\cite{Peskin:1991sw}
%\bibitem{Peskin:1991sw}
  M.~E.~Peskin and T.~Takeuchi,
  %``Estimation of oblique electroweak corrections,''
  Phys.\ Rev.\ D {\bf 46}, 381 (1992).
 % doi:10.1103/PhysRevD.46.381
  %%CITATION = doi:10.1103/PhysRevD.46.381;%%
  %1641 citations counted in INSPIRE as of 03 Dec 2015

%\cite{Hsieh:2010zr}
\bibitem{Hsieh:2010zr}
  K.~Hsieh, K.~Schmitz, J.~H.~Yu and C.-P.~Yuan,
  %``Global Analysis of General SU(2) x SU(2) x U(1) Models with Precision Data,''
  Phys.\ Rev.\ D {\bf 82}, 035011 (2010)
 % doi:10.1103/PhysRevD.82.035011
  [arXiv:1003.3482 [hep-ph]].
  %%CITATION = doi:10.1103/PhysRevD.82.035011;%%
  %48 citations counted in INSPIRE as of 03 Dec 2015

%\cite{Deppisch:2014qpa}
\bibitem{Deppisch:2014qpa}
  F.~F.~Deppisch, T.~E.~Gonzalo, S.~Patra, N.~Sahu and U.~Sarkar,
  %``Signal of Right-Handed Charged Gauge Bosons at the LHC?,''
  Phys.\ Rev.\ D {\bf 90}, no. 5, 053014 (2014)
  [arXiv:1407.5384 [hep-ph]];
  %%CITATION = ARXIV:1407.5384;%%
  %38 citations counted in INSPIRE as of 17 sept. 2015
%\cite{Heikinheimo:2014tba}
%\bibitem{Heikinheimo:2014tba}
  M.~Heikinheimo, M.~Raidal and C.~Spethmann,
  %``Testing Right-Handed Currents at the LHC,''
  Eur.\ Phys.\ J.\ C {\bf 74}, no. 10, 3107 (2014)
  [arXiv:1407.6908 [hep-ph]];
  %%CITATION = ARXIV:1407.6908;%%
  %36 citations counted in INSPIRE as of 17 sept. 2015
%\cite{Deppisch:2014zta}
%\bibitem{Deppisch:2014zta}
  F.~F.~Deppisch, T.~E.~Gonzalo, S.~Patra, N.~Sahu and U.~Sarkar,
  %``Double beta decay, lepton flavor violation, and collider signatures of left-right symmetric models with spontaneous $D$-parity breaking,''
  Phys.\ Rev.\ D {\bf 91}, no. 1, 015018 (2015)
  [arXiv:1410.6427 [hep-ph]].
  %%CITATION = ARXIV:1410.6427;%%
  %17 citations counted in INSPIRE as of 17 sept. 2015

%\cite{Aad:2015xaa}
\bibitem{Aad:2015xaa}
  G.~Aad {\it et al.} [ATLAS Collaboration],
  %``Search for heavy Majorana neutrinos with the ATLAS detector in pp collisions at $ \sqrt{s}=8 $ TeV,''
  JHEP {\bf 1507}, 162 (2015)
  [arXiv:1506.06020 [hep-ex]].
  %%CITATION = ARXIV:1506.06020;%%
  %6 citations counted in INSPIRE as of 18 sept. 2015

%\cite{Mohapatra:1986aw}
\bibitem{Mohapatra:1986aw}
  R.~N.~Mohapatra,
  %``Mechanism for Understanding Small Neutrino Mass in Superstring Theories,''
  Phys.\ Rev.\ Lett.\  {\bf 56}, 561 (1986);
  %%CITATION = PRLTA,56,561;%%
  %266 citations counted in INSPIRE as of 18 sept. 2015
%\cite{Mohapatra:1986bd}
%\bibitem{Mohapatra:1986bd}
  R.~N.~Mohapatra and J.~W.~F.~Valle,
  %``Neutrino Mass and Baryon Number Nonconservation in Superstring Models,''
  Phys.\ Rev.\ D {\bf 34}, 1642 (1986).
  %%CITATION = PHRVA,D34,1642;%%
  %665 citations counted in INSPIRE as of 18 sept. 2015

%\cite{Barr:2003nn}
\bibitem{Barr:2003nn}
  S.~M.~Barr,
  %``A Different seesaw formula for neutrino masses,''
  Phys.\ Rev.\ Lett.\  {\bf 92}, 101601 (2004)
  [hep-ph/0309152];
  %%CITATION = HEP-PH/0309152;%%
  %75 citations counted in INSPIRE as of 18 sept. 2015
%\cite{Malinsky:2005bi}
%\bibitem{Malinsky:2005bi}
  M.~Malinsky, J.~C.~Romao and J.~W.~F.~Valle,
  %``Novel supersymmetric SO(10) seesaw mechanism,''
  Phys.\ Rev.\ Lett.\  {\bf 95}, 161801 (2005)
  [hep-ph/0506296];
  %%CITATION = HEP-PH/0506296;%%
  %136 citations counted in INSPIRE as of 18 sept. 2015
%\cite{Fukuyama:2005us}
%\bibitem{Fukuyama:2005us}
  T.~Fukuyama, T.~Kikuchi and T.~Osaka,
  %``Non-thermal leptogenesis and a prediction of inflaton mass in a supersymmetric SO(10) model,''
  JCAP {\bf 0506}, 005 (2005)
  [hep-ph/0503201].
  %%CITATION = HEP-PH/0503201;%%
  %17 citations counted in INSPIRE as of 18 sept. 2015


%\cite{Liew:2015osa}
\bibitem{Liew:2015osa}
  S.~P.~Liew and S.~Shirai,
  %``Testing ATLAS Diboson Excess with Dark Matter Searches at LHC,''
  arXiv:1507.08273 [hep-ph].
  %%CITATION = ARXIV:1507.08273;%%
  %11 citations counted in INSPIRE as of 03 Dec 2015

%\cite{Cohen:2013ama}
\bibitem{Cohen:2013ama}
  T.~Cohen, M.~Lisanti, A.~Pierce and T.~R.~Slatyer,
  %``Wino Dark Matter Under Siege,''
  JCAP {\bf 1310}, 061 (2013)
%  doi:10.1088/1475-7516/2013/10/061
  [arXiv:1307.4082];
  %%CITATION = doi:10.1088/1475-7516/2013/10/061;%%
  %97 citations counted in INSPIRE as of 03 Dec 2015
%\cite{Fan:2013faa}
%\bibitem{Fan:2013faa}
  J.~Fan and M.~Reece,
  %``In Wino Veritas? Indirect Searches Shed Light on Neutralino Dark Matter,''
  JHEP {\bf 1310}, 124 (2013)
 % doi:10.1007/JHEP10(2013)124
  [arXiv:1307.4400 [hep-ph]];
  %%CITATION = doi:10.1007/JHEP10(2013)124;%%
  %97 citations counted in INSPIRE as of 03 Dec 2015
%\cite{Hryczuk:2014hpa}
%\bibitem{Hryczuk:2014hpa}
  A.~Hryczuk, I.~Cholis, R.~Iengo, M.~Tavakoli and P.~Ullio,
  %``Indirect Detection Analysis: Wino Dark Matter Case Study,''
  JCAP {\bf 1407}, 031 (2014)
 % doi:10.1088/1475-7516/2014/07/031
  [arXiv:1401.6212 [astro-ph.HE]];
  %%CITATION = doi:10.1088/1475-7516/2014/07/031;%%
  %46 citations counted in INSPIRE as of 03 Dec 2015
%\cite{Bhattacherjee:2014dya}
%\bibitem{Bhattacherjee:2014dya}
  B.~Bhattacherjee, M.~Ibe, K.~Ichikawa, S.~Matsumoto and K.~Nishiyama,
  %``Wino Dark Matter and Future dSph Observations,''
  JHEP {\bf 1407}, 080 (2014)
 % doi:10.1007/JHEP07(2014)080
  [arXiv:1405.4914 [hep-ph]].
  %%CITATION = doi:10.1007/JHEP07(2014)080;%%
  %22 citations counted in INSPIRE as of 03 Dec 2015

%\cite{Hisano:2015rsa}
\bibitem{Hisano:2015rsa}
  J.~Hisano, K.~Ishiwata and N.~Nagata,
  %``QCD Effects on Direct Detection of Wino Dark Matter,''
  JHEP {\bf 1506}, 097 (2015)
%  doi:10.1007/JHEP06(2015)097
  [arXiv:1504.00915 [hep-ph]].
  %%CITATION = doi:10.1007/JHEP06(2015)097;%%
  %14 citations counted in INSPIRE as of 03 Dec 2015

%\cite{Khachatryan:2015dcf}
\bibitem{Khachatryan:2015dcf}
  V.~Khachatryan {\it et al.} [CMS Collaboration],
  %``Search for narrow resonances decaying to dijets in proton-proton collisions at sqrt(s) = 13 TeV,''
  arXiv:1512.01224 [hep-ex].
  %%CITATION = ARXIV:1512.01224;%%
  
  %\cite{ATLAS:2015nsi}
\bibitem{ATLAS:2015nsi} 
  [ATLAS Collaboration],
  %``Search for New Phenomena in Dijet Mass and Angular Distributions with the ATLAS Detector at $\sqrt{s}$ = 13 TeV,''
  arXiv:1512.01530 [hep-ex].
  %%CITATION = ARXIV:1512.01530;%%

\end{thebibliography}
\end{document}